
\input harvmac
\input epsf
\input amssym
%
\noblackbox


\def\coeff#1#2{\relax{\textstyle {#1 \over #2}}\displaystyle}

\def\cC{{\cal C}}

 \def\cM{{\cal M}}
 
 \def\cQ{{\cal Q}}
\def\cR{{\cal R}} \def\cS{{\cal S}}

\def\ZZ{\Bbb{Z}}

\def\IR{\Bbb{R}}
%


\lref\LuninJY{
  O.~Lunin and S.~D.~Mathur,
``AdS/CFT duality and the black hole information paradox,''
  Nucl.\ Phys.\  B {\bf 623}, 342 (2002)
  [arXiv:hep-th/0109154].
}

\lref\LuninQF{
  O.~Lunin and S.~D.~Mathur,
``Statistical interpretation of Bekenstein entropy for systems with a
stretched horizon,''
  Phys.\ Rev.\ Lett.\  {\bf 88}, 211303 (2002)
  [arXiv:hep-th/0202072].
}

\lref\LuninIZ{
  O.~Lunin, J.~M.~Maldacena and L.~Maoz,
``Gravity solutions for the D1-D5 system with angular momentum,''
  arXiv:hep-th/0212210.
}

\lref\LuninDT{
  O.~Lunin and S.~D.~Mathur,
``The slowly rotating near extremal D1-D5 system as a 'hot tube',''
  Nucl.\ Phys.\  B {\bf 615}, 285 (2001)
  [arXiv:hep-th/0107113].
}

\lref\AldayND{
  L.~F.~Alday, J.~de Boer and I.~Messamah,
``The gravitational description of coarse grained microstates,''
  JHEP {\bf 0612}, 063 (2006)
  [arXiv:hep-th/0607222].
}

\lref\DonosVS{
  A.~Donos and A.~Jevicki,
``Dynamics of chiral primaries in AdS(3) x S**3 x T**4,''
  Phys.\ Rev.\  D {\bf 73}, 085010 (2006)
  [arXiv:hep-th/0512017].
}

\lref\AldayXJ{
  L.~F.~Alday, J.~de Boer and I.~Messamah,
``What is the dual of a dipole?,''
  Nucl.\ Phys.\  B {\bf 746}, 29 (2006)
  [arXiv:hep-th/0511246].
}

\lref\GiustoAG{
  S.~Giusto, S.~D.~Mathur and Y.~K.~Srivastava,
``Dynamics of supertubes,''
  Nucl.\ Phys.\  B {\bf 754}, 233 (2006)
  [arXiv:hep-th/0510235].
}

\lref\TaylorDB{
  M.~Taylor,
``General 2 charge geometries,''
  JHEP {\bf 0603}, 009 (2006)
  [arXiv:hep-th/0507223].
}

\lref\SkenderisAH{
  K.~Skenderis and M.~Taylor,
``Fuzzball solutions and D1-D5 microstates,''
  Phys.\ Rev.\ Lett.\  {\bf 98}, 071601 (2007)
  [arXiv:hep-th/0609154].
}

\lref\IizukaUV{
  N.~Iizuka and M.~Shigemori,
``A note on D1-D5-J system and 5D small black ring,''
  JHEP {\bf 0508}, 100 (2005)
  [arXiv:hep-th/0506215].
}

\lref\BoniSF{
  M.~Boni and P.~J.~Silva,
``Revisiting the D1/D5 system or bubbling in AdS(3),''
  JHEP {\bf 0510}, 070 (2005)
  [arXiv:hep-th/0506085].
}

\lref\MartelliXQ{
  D.~Martelli and J.~F.~Morales,
``Bubbling AdS(3),''
  JHEP {\bf 0502}, 048 (2005)
  [arXiv:hep-th/0412136].
}

\lref\SrivastavaPA{
  Y.~K.~Srivastava,
``Perturbations of supertube in KK monopole background,''
  arXiv:hep-th/0611320.
}

\lref\KanitscheiderWQ{
  I.~Kanitscheider, K.~Skenderis and M.~Taylor,
``Fuzzballs with internal excitations,''
  arXiv:0704.0690 [hep-th].
}

\lref\KanitscheiderZF{
  I.~Kanitscheider, K.~Skenderis and M.~Taylor,
``Holographic anatomy of fuzzballs,''
  JHEP {\bf 0704}, 023 (2007)
  [arXiv:hep-th/0611171].
}

\lref\SkenderisYB{
  K.~Skenderis and M.~Taylor,
``Anatomy of bubbling solutions,''
  arXiv:0706.0216 [hep-th].
}

\lref\BenaVA{
  I.~Bena and N.~P.~Warner,
``Bubbling supertubes and foaming black holes,''
  Phys.\ Rev.\  D {\bf 74}, 066001 (2006)
  [arXiv:hep-th/0505166].
}

\lref\BerglundVB{
  P.~Berglund, E.~G.~Gimon and T.~S.~Levi,
``Supergravity microstates for BPS black holes and black rings,''
  JHEP {\bf 0606}, 007 (2006)
  [arXiv:hep-th/0505167].
}

\lref\BenaIS{
  I.~Bena, C.~W.~Wang and N.~P.~Warner,
``The foaming three-charge black hole,''
  arXiv:hep-th/0604110.
}

\lref\BenaKB{
  I.~Bena, C.~W.~Wang and N.~P.~Warner,
``Mergers and typical black hole microstates,''
  JHEP {\bf 0611}, 042 (2006)
  [arXiv:hep-th/0608217].
}

\lref\BenaKG{
  I.~Bena and N.~P.~Warner,
``Black holes, black rings and their microstates,''
  arXiv:hep-th/0701216.
}

\lref\BenaZY{
  I.~Bena, C.~W.~Wang and N.~P.~Warner,
``Sliding rings and spinning holes,''
  JHEP {\bf 0605}, 075 (2006)
  [arXiv:hep-th/0512157].
}

\lref\GiustoKJ{
  S.~Giusto and S.~D.~Mathur,
``Geometry of D1-D5-P bound states,''
  Nucl.\ Phys.\  B {\bf 729}, 203 (2005)
  [arXiv:hep-th/0409067].
}

\lref\DenefNB{
  F.~Denef,
``Supergravity flows and D-brane stability,''
  JHEP {\bf 0008}, 050 (2000)
  [arXiv:hep-th/0005049].
}

\lref\DenefRU{
  F.~Denef,
``Quantum quivers and Hall/hole halos,''
  JHEP {\bf 0210}, 023 (2002)
  [arXiv:hep-th/0206072].
}

\lref\BatesVX{
  B.~Bates and F.~Denef,
``Exact solutions for supersymmetric stationary black hole composites,''
  arXiv:hep-th/0304094.
}

\lref\FreedmanTZ{
  D.~Z.~Freedman, S.~D.~Mathur, A.~Matusis and L.~Rastelli,
``Correlation functions in the CFT($d$)/AdS($d+1$) correspondence,''
  Nucl.\ Phys.\  B {\bf 546}, 96 (1999)
  [arXiv:hep-th/9804058].
}

\lref\GiustoID{
  S.~Giusto, S.~D.~Mathur and A.~Saxena,
  ``Dual geometries for a set of 3-charge microstates,''
  Nucl.\ Phys.\  B {\bf 701}, 357 (2004)
  [arXiv:hep-th/0405017].
}

\lref\GiustoIP{
  S.~Giusto, S.~D.~Mathur and A.~Saxena,
  ``3-charge geometries and their CFT duals,''
  Nucl.\ Phys.\  B {\bf 710}, 425 (2005)
  [arXiv:hep-th/0406103].
}

\lref\FordYB{
  J.~Ford, S.~Giusto and A.~Saxena,
  ``A class of BPS time-dependent 3-charge microstates from spectral flow,''
  arXiv:hep-th/0612227.
}

\lref\MathurHJ{
  S.~D.~Mathur, A.~Saxena and Y.~K.~Srivastava,
  ``Constructing 'hair' for the three charge hole,''
  Nucl.\ Phys.\  B {\bf 680}, 415 (2004)
  [arXiv:hep-th/0311092].
}

\lref\BenaWT{
  I.~Bena and P.~Kraus,
  ``Three charge supertubes and black hole hair,''
  Phys.\ Rev.\  D {\bf 70}, 046003 (2004)
  [arXiv:hep-th/0402144].
}

\lref\LuninUU{
  O.~Lunin,
  ``Adding momentum to D1-D5 system,''
  JHEP {\bf 0404}, 054 (2004)
  [arXiv:hep-th/0404006].
}

\lref\BalasubramanianGI{
  V.~Balasubramanian, E.~G.~Gimon and T.~S.~Levi,
  ``Four Dimensional Black Hole Microstates: From D-branes to Spacetime Foam,''
  arXiv:hep-th/0606118.
}
\lref\BenaAY{
  I.~Bena and P.~Kraus,
  ``Microstates of the D1-D5-KK system,''
  Phys.\ Rev.\  D {\bf 72}, 025007 (2005)
  [arXiv:hep-th/0503053].
}

\lref\SaxenaUK{
  A.~Saxena, G.~Potvin, S.~Giusto and A.~W.~Peet,
  ``Smooth geometries with four charges in four dimensions,''
  JHEP {\bf 0604}, 010 (2006)
  [arXiv:hep-th/0509214].
}

\lref\DenefYT{
  F.~Denef, D.~Gaiotto, A.~Strominger, D.~Van den Bleeken and X.~Yin,
  ``Black hole deconstruction,''
  arXiv:hep-th/0703252.
}


\lref\GimonPS{
  E.~G.~Gimon, T.~S.~Levi and S.~F.~Ross,
  ``Geometry of non-supersymmetric three-charge bound states,''
  arXiv:0705.1238 [hep-th].
}

\lref\JejjalaYU{
  V.~Jejjala, O.~Madden, S.~F.~Ross and G.~Titchener,
  ``Non-supersymmetric smooth geometries and D1-D5-P bound states,''
  Phys.\ Rev.\  D {\bf 71}, 124030 (2005)
  [arXiv:hep-th/0504181].
}

\lref\samirreview{
  S.~D.~Mathur,
  ``The quantum structure of black holes,''
  Class.\ Quant.\ Grav.\  {\bf 23}, R115 (2006)
  [arXiv:hep-th/0510180].
  S.~D.~Mathur,
  ``The fuzzball proposal for black holes: An elementary review,''
  Fortsch.\ Phys.\  {\bf 53}, 793 (2005)
  [arXiv:hep-th/0502050].
}

\lref\moreGH{
  M.~C.~N.~Cheng,
  ``More bubbling solutions,''
  JHEP {\bf 0703}, 070 (2007)
  [arXiv:hep-th/0611156].
}

\lref\elvangtwo{
  H.~Elvang, R.~Emparan, D.~Mateos and H.~S.~Reall,
  ``Supersymmetric black rings and three-charge supertubes,''
  Phys.\ Rev.\  D {\bf 71}, 024033 (2005)
  [arXiv:hep-th/0408120].
}

\lref\ElvangRT{
  H.~Elvang, R.~Emparan, D.~Mateos and H.~S.~Reall,
  ``A supersymmetric black ring,''
  Phys.\ Rev.\ Lett.\  {\bf 93}, 211302 (2004)
  [arXiv:hep-th/0407065].
}

\lref\BenaDE{
  I.~Bena and N.~P.~Warner,
  ``One ring to rule them all ... and in the darkness bind them?,''
  Adv.\ Theor.\ Math.\ Phys.\  {\bf 9}, 667 (2005)
  [arXiv:hep-th/0408106].
}
\lref\BenaWV{
  I.~Bena,
  ``Splitting hairs of the three charge black hole,''
  Phys.\ Rev.\  D {\bf 70}, 105018 (2004)
  [arXiv:hep-th/0404073].
}

\lref\GauntlettQY{
  J.~P.~Gauntlett and J.~B.~Gutowski,
  ``General concentric black rings,''
  Phys.\ Rev.\  D {\bf 71}, 045002 (2005)
  [arXiv:hep-th/0408122].
}
\lref\GauntlettWH{
  J.~P.~Gauntlett and J.~B.~Gutowski,
  ``Concentric black rings,''
  Phys.\ Rev.\  D {\bf 71}, 025013 (2005)
  [arXiv:hep-th/0408010].
}

\lref\GauntlettNW{
  J.~P.~Gauntlett, J.~B.~Gutowski, C.~M.~Hull, S.~Pakis and H.~S.~Reall,
  ``All supersymmetric solutions of minimal supergravity in five dimensions,''
  Class.\ Quant.\ Grav.\  {\bf 20}, 4587 (2003)
  [arXiv:hep-th/0209114].
}

\lref\entropy{
  I.~Bena and P.~Kraus,
  ``Microscopic description of black rings in AdS/CFT,''
  JHEP {\bf 0412}, 070 (2004)
  [arXiv:hep-th/0408186].
}

\lref\UnruhIC{
  W.~G.~Unruh and R.~M.~Wald,
`Acceleration Radiation And Generalized Second Law Of Thermodynamics,''
  Phys.\ Rev.\  D {\bf 25}, 942 (1982).
}

\lref\UnruhIR{
  W.~G.~Unruh and R.~M.~Wald,
``Entropy Bounds, Acceleration Radiation, And The Generalized Second Law,''
  Phys.\ Rev.\  D {\bf 27}, 2271 (1983).
}

\lref\MaldacenaUZ{
  J.~M.~Maldacena, J.~Michelson and A.~Strominger,
  ``Anti-de Sitter fragmentation,''
  JHEP {\bf 9902}, 011 (1999)
  [arXiv:hep-th/9812073].
}

\lref\MarolfAY{
  D.~Marolf and R.~Sorkin,
  ``Perfect mirrors and the self-accelerating box paradox,''
  Phys.\ Rev.\  D {\bf 66}, 104004 (2002)
  [arXiv:hep-th/0201255].
}

\lref\eric{ E.~G.~Gimon and  T.~S.~Levi,
  ``Black Ring Deconstruction,''
  arXiv:0706.3394 [hep-th].
}

\lref\DenefVG{
  F.~Denef and G.~W.~Moore,
  ``Split states, entropy enigmas, holes and halos,''
  arXiv:hep-th/0702146.
}



\Title{
\vbox{
\hbox{\baselineskip12pt \vbox{\hbox{SPhT-T07/075} }
}}}
{\vbox{\vskip -1.5cm
\centerline{\hbox{Plumbing the Abyss: Black Ring Microstates}}
}}
\vskip -.3cm
\centerline{Iosif~Bena${}^{(1)}$,  Chih-Wei Wang${}^{(2)}$ and
Nicholas P.\ Warner${}^{(2)}$}

\bigskip
\centerline{$^{(1)}$ \it Service de Physique Th\'eorique, }
\centerline{\it CEA Saclay, 91191 Gif sur Yvette, France}
\bigskip
\centerline{{${}^{(2)}$\it Department of Physics and Astronomy,
University of Southern California}} \centerline{{\it Los Angeles,
CA 90089-0484, USA}}
\medskip
\centerline{{\rm iosif.bena@cea.fr, chihweiw@usc.edu,  warner@usc.edu} }
\bigskip
\bigskip

We construct the first smooth, horizonless ``microstate geometries'' that have
the same charges, dipole charges and angular momenta as a BPS black ring whose
horizon is macroscopic.  These solutions have exactly the same geometry as
black rings, except that the usual infinite throat is smoothly capped off at a very large
depth. If the solutions preserve a $U(1) \times U(1)$ isometry, then
this depth is limited by flux quantization but if this
symmetry is broken then the throat can be made arbitrarily
deep by tuning classical, geometric moduli. Interpreting these ``abysses''
(smooth microstate geometries of arbitrary depth) from the point of view of the $AdS$-CFT
correspondence suggests two remarkable alternatives: either stringy effects can eliminate
very large regions of a smooth low-curvature supergravity solution, or the D1-D5-P CFT has
quantum critical points. The existence of solutions whose depth depends on moduli also
enables us to define ``entropy elevators,'' and these provide a new tool for studying
the entropy of BPS and near-BPS black holes.

\vskip .3in
\Date{\sl {June, 2007}}

\vfill\eject

\newsec{Introduction}

While it is still uncertain whether it is a misnomer, the term
``microstate geometries'' has come to represent smooth, horizonless
geometries that represent good string backgrounds, are asymptotic to
$\IR^{n,1} \times \cC$, where $\cC$ is compact, and have the
same charges as black holes or black rings.
Given that such
geometries exist at all (and apparently in large numbers) and that
 they provide semi-classical descriptions of microstates of
black holes and black rings, it is important to investigate their
physics. It is also very tempting to conjecture that these
geometries could account for the entropy of black objects.

This conjecture leads to a host of interesting physical
questions.  Do ``microstate geometries'' represent extremely special,
coherent states of the stringy black hole or
can microstate geometries represent the ``typical'' microstates of a
black hole and contribute significantly to its entropy?  Are there
enough classical microstate geometries to account for the black hole
entropy? Is there a semi-classical quantization of these microstate
geometries that leads to the correct black-hole entropy, at least to leading
order?  If so, can the typical microstate geometries be described in
supergravity, or are
they necessarily stringy?  Many of these and related issues have been thoroughly
analyzed for the two-charge system (see
\refs{\LuninJY\LuninQF\LuninIZ\LuninDT\AldayND\DonosVS
\AldayXJ\GiustoAG\TaylorDB\SkenderisAH\KanitscheiderZF{--}\KanitscheiderWQ}
for an incomplete list of relevant papers,
and \samirreview\ for a review), and there is now
very significant progress on the three-charge system (see \refs{\BenaWT\GiustoID\LuninUU\GiustoIP\MathurHJ\GiustoKJ\BenaVA\BerglundVB\BenaIS\BenaKB\FordYB{--}\moreGH} and \BenaKG\ for a review), as well as for the four-dimensional \refs{\BenaAY\SaxenaUK\BalasubramanianGI\SrivastavaPA{--}\DenefYT} and non-BPS black-hole microstates \refs{\JejjalaYU,\GimonPS}.

One of the key steps in understanding three-charge microstate
geometries was the realization that they are bubbling geometries, which
come from the geometric transition of three-charge brane
configurations \refs{\GiustoKJ,\BenaVA,\BerglundVB}.  Such a
transition replaces the spatial $\IR^4$ that is used to construct the black-hole and
black-ring solutions by a topologically non-trivial
hyper-K\"ahler manifold. The singular sources that define the
original black holes or black rings are replaced by smooth topological
fluxes threading non-trivial cycles.

For reasons of computational simplicity, the metrics on the hyper-K\"ahler base were usually
chosen to be Gibbons-Hawking (GH) metrics and microstate solutions constructed using these metrics have been
analyzed extensively \refs{\GiustoKJ,\BenaVA,\BerglundVB,\moreGH}.  In particular,
the asymptotic charges and angular momenta were computed and it was found that generic
distributions of fluxes lead to microstate geometries whose charges
correspond to maximally-spinning classical black holes or black rings (of zero horizon
area) \BenaIS.  We typically refer to such microstates geometries as ``zero-entropy microstate
geometries\foot{One should, of course, remember that a microstate
geometry is horizonless and smooth and necessarily has zero entropy.
The phrase ``zero-entropy microstate geometries'' is meant to emphasize the fact that the
corresponding classical black object with the same charges and angular
momenta also has vanishing entropy.}.''

It is not known generically how to evade the restriction to zero-entropy microstates
without introducing closed time-like curves (CTC's). At present the only systematic way of
doing this is to use mergers of zero-entropy microstates to obtain
microstates of objects with non-zero entropy \BenaZY.
This technique was used in \BenaKB\ to construct the first microstate solutions that have the same
charges and angular momenta as a three-charge black hole of classically large horizon area, that is, a ``true'' black hole from the perspective of classical general relativity. In this paper we will also use both mergers of two zero-entropy black-ring microstates, as well as other methods, to obtain
the first microstates of black rings with a classical horizon area.
Note that because of the infinite  non-uniqueness of BPS black rings
\refs{\BenaDE,\elvangtwo,\GauntlettQY}, a solution with black ring charges and angular momenta is not
necessarily a black ring microstate. A microstate has (by definition) {\it all} the macroscopic features of the object it describes, and for black ring this includes not only the charges and angular momenta, but also the dipole charges \refs{\BenaWV,\ElvangRT,\BenaDE,\elvangtwo,\GauntlettQY}.

One of the interesting, and probably defining features of the microstates geometries of
true black holes is that they are scaling solutions  or ``deep microstates.''
That is, the ``bubbles,'' or non-trivial topological cycles scale into a vanishingly small
region in the GH base metric (while preserving the relative sizes of the cycles).
In the physical, space-time geometry this corresponds to the cycles descending
deeply into a black-hole-like throat.  Thus the bubbled black-hole geometries
look like a regular black hole, except that their throat is capped off by regular
geometry deep down the throat.  In \BenaKB\ it was shown that, at least
for $U(1) \times U(1)$ invariant geometries, the depth at which the capping-off
occurs is set by the size of the smallest quantum of flux that one can place
upon a single bubble.  Moreover, it was shown that small, non-BPS fluctuations in
the region of the cap have an energy (as measured from infinity)
that matches the expected mass gap of typical states in the underlying
D1-D5 conformal field theory (CFT).  Thus these ``deep microstates'' should
be interpreted as the holographic duals the long effective strings of
the D1-D5 CFT.

Our purpose here is to study deep microstate geometries in more detail, focussing
first on the mergers of bubbled supertube geometries that lead to microstate
geometries of BPS black rings with a classically large horizon area (that is, ``true'' black rings).
After reviewing some basic properties
of bubbled geometries in Section 2, we begin in Section 3 by considering microstate
geometries  corresponding to a pair of supertubes and study the axially
symmetric ($U(1) \times U(1)$ invariant) mergers that lead to  microstate
geometries of ``true'' black rings. In Section 4 we
find that this merger results in scaling, or deep, geometries where the non-trivial cycles descend deeply into
the $AdS$ throat of what looks like a classical black ring.  The non-trivial topology, once
again, smoothly caps off this throat at a depth that is set by the quantum of
flux on an individual cycle.  We thus obtain the analogs for black rings of the results
that were found for black holes in \BenaKB.

The reader who is not keen on the technical details of the construction, and only interested in the physics of smooth microstate solutions of arbitrary depth can skip directly to Section 5, where we construct the first
such example by considering a scaling solution that is no longer axi-symmetric.  This solution has the surprising feature that the depth of the throat can be controlled by a modulus and can be made arbitrarily large by fine-tuning this modulus\foot{The solution we study is a five-dimensional, smooth solution
that, from a four-dimensional perspective, corresponds to a ``closed quiver'' of
D6 and anti-D6 branes \refs{\DenefNB\DenefRU\BatesVX{--}\DenefVG}.}.  During this process
the solution remains {\it completely smooth}.  We will refer to such throats of arbitrary depth as
{\it abysses}. The existence of abysses suggests that breaking the $U(1) \times
U(1)$ invariance allows the construction of smooth horizonless geometries whose holographic duals in
the CFT exhibit mass gaps and spectra with energy gaps that are arbitrarily
small.  We discuss the interesting physics implied by the existence of these new solutions is Section 6.
The properties of these solutions also suggest they can be used as ``entropy elevators,'' which could account for the entropy both of BPS and of non-BPS black holes. This idea is developed in Section 7.

{\bf Note Added:} The day before this paper was submitted to the arXiv, the preprint \eric\ appeared,
in which a
horizonless three-charge scaling configuration with black ring charges is also constructed. From a four-dimensional perspective this configuration contains D6 and $\overline {\rm D6}$ branes, like the solutions we construct here, but also contains D0 branes. Hence, the five-dimensional lift of this solution, as well as of the solutions analyzed in \DenefYT, has a naked singularity corresponding to the D0 branes.  Therefore, the solutions in \refs{\DenefYT,\eric} are useful for understanding and counting black hole microstates in the intermediate regime of parameters where the D4 branes affect the geometry and the D0's are considered as probes. However, because of the naked singularity, their fate in the regime of parameters where black holes and black rings have macroscopic horizons is unknown. In contrast, solutions with only D6 and $\overline {\rm D6}$ remain smooth in five dimensions, and hence give a valid description of microstates of black holes and black rings in the same region of the moduli space where the classical black holes and black rings also exist.


\newsec{Bubbled Geometries}

Before focussing on bubbled ring geometries, it is worthwhile reviewing
some of the basics of bubbled geometries in general.  For more details, see
\refs{\BenaVA\BerglundVB\BenaIS{--}\BenaKB,\BenaKG}.
First recall that the four-dimensional base metric has Gibbons-Hawking (GH) form:
\eqn\GHmetric{ds_4^2 ~=~  V^{-1} \, \big( d\psi + \vec{A} \cdot
d\vec{y}\big)^2  ~+~ V\, (d\vec{y}\cdot d\vec{y} )\,,}
where $\vec y \in \IR^3$ and
\eqn\Vform{V ~=~   \sum_{j=1}^N \,  {q_j  \over r_j} \,, \qquad
\vec \nabla \times \vec A ~=~ \vec \nabla V \,,}
with  $r_j \equiv |\vec{y}-\vec{y}^{(j)}|$.   In order for the GH metric
to be regular, one must take $q_j \in \ZZ$ and for the metric to be asymptotic
to that of flat $\IR^4$ one must also impose
\eqn\qzero{q_0 ~\equiv~ \sum_{j=1}^N \, q_j  ~=~ 1\,.}
The fluxes through the non-trivial two-cycles in this geometry are determined
by harmonic functions:
\eqn\KIdefn{K^I ~\equiv~   \sum_{j=1}^N \, {k_j^I \over r_j} \,,}
and by the flux parameters,  $k_j^I $, in particular.   There is a gauge
equivalence $K^I \to K^I + c^I V$, or $k_j^I \to  k_j^I + c^I q_j$ for any constant, $c^I$.
It is therefore useful to define the gauge invariant flux parameters:
\eqn\ktilde{\tilde  k^I_j ~\equiv~ k^I_j ~-~    q_j\, N  \,  k_0^I  \,,
\qquad {\rm with} \qquad k_0^I ~\equiv~{1 \over N} \, \sum_{j=1}^N k_j^I\,.}
Following \refs{\BenaVA,\BerglundVB,\BenaKB}, the charges and
angular momenta  of a bubbled solution are can be obtained from   the positions,
$\vec y^{(j)} $, of the  GH points via:
\eqn\QIchg{Q_I ~=~ -2 \, C_{IJK} \, \sum_{j=1}^N \, q_j^{-1} \,
\tilde  k^J_j \, \tilde  k^K_j\,,}
\eqn\Jright{ J_R ~\equiv~ J_1 + J_2 ~=~ \coeff{4}{3}\, \, C_{IJK} \, \sum_{j=1}^N q_j^{-2} \,
\tilde  k^I_j \, \tilde  k^J_j \,  \tilde  k^K_j  \,,}
\eqn\Jleft{ J_L ~\equiv~ J_1 - J_2 ~=~ 8 \,\big| \vec D\big|  \,,}
where $N$ is the number of GH points and
\eqn\dipoles{\vec D_j ~\equiv~  \, \sum_I  \, \tilde k_j^I \, \vec y^{(j)} \,,
\qquad \vec D ~\equiv~ \sum_{j=1}^N \, \vec D_j \,.}

It is convenient to define:
\eqn\Pijdefn{P_{ij} ~\equiv~   \coeff{1}{6}\, C_{IJK} \, \Pi^{(I)}_{ij}\, \Pi^{(J)}_{ij}\,
\Pi^{(K)}_{ij}  \,}
and
\eqn\angmomflux{\vec J_{L\, ij} ~\equiv ~ -  8 \,q_i \, q_j  \,
P_{ij}  \, \hat y_{ij} \,, \qquad{\rm where} \qquad \hat y_{ij} ~\equiv ~
 {(\vec y^{(i)} - \vec y^{(j)}) \over \big|\vec y^{(i)} - \vec y^{(j)}\big| } \,. }
One then has  \refs{\BerglundVB,\BenaKB}:
\eqn\JLsimp{\eqalign{\vec J_L ~=~  \sum_{{\scriptstyle i, j=1} \atop
{\scriptstyle j \ne i}}^N \,  \vec J_{L\, ij}   \,,  }}
and if the GH points are all co-linear then we may take
$\hat y_{ij} = \pm 1$ and \JLsimp\  reduces to a sum over $\pm P_{ij}$.

Finally, to eliminate CTC's near the GH points, this configuration must
satisfy the bubble equations:
\eqn\BubbleEqns{
 \sum_{{\scriptstyle j=1} \atop {\scriptstyle j \ne i}}^N \,
 q_i \, q_j  \,   {P_{ij} \over r_{ij} } ~=~ - \sum_{I=1}^3  \tilde k^I_i   \,, }
where $r_{ij} = |\vec y^{(i)} - \vec y^{(j)} | $

In this paper, we specifically wish to consider the situation where the bubbled
geometry looks, at large scales, like a supertube or black ring.  We therefore
wish to take $q_1 =+ 1$ and locate this GH point at $\vec y^{(1)} = 0$ and
will then assume that all the remaining GH points cluster a some distance,
$\rho$ from the origin.  More specifically, we will typically assume that
\eqn\bubbring{r_{1j} ~\approx~  \rho \,, \qquad   r_{ij} ~\ll~ \rho \,, \qquad
i,j =2,\dots,N \,.}
In this limit, the first bubble equation yields the approximate ring radius:
\eqn\ringrad{\rho ~\approx~ -\bigg[\sum_{I=1}^3  \tilde k^I_1 \bigg]^{-1} \,
\sum_{j=2}^N \, q_j  \,    P_{1j}     \,.}
%

\newsec{The axi-symmetric merger of two bubbled supertubes}

\subsec{The layout and physical parameters}

We consider the simplest possible pair of bubbled rings in which each ring
is bubbled by identical pairs of GH points of charges,  $-Q$ and $+Q$.
Thus the configuration will have:
\eqn\fivechgs{q_1 ~=~  +1\,, \qquad q_2 ~=~  -Q \,, \qquad q_3 ~=~  +Q\,, \qquad
q_4 ~=~  -Q \,, \qquad q_5 ~=~  +Q\,,}
and  we will denote the various  distances by
\eqn\distparams{ \rho ~\equiv~  r_{12}\,, \qquad \sigma ~\equiv~  r_{34} \,, \qquad
\Delta_1 ~\equiv~  r_{23} \,, \qquad \Delta_2 ~\equiv~  r_{45} \,.}
This layout is depicted in Fig.~1.
The $\Delta_j$ will represent the bubbled ring widths, and
in the classical  limit, $Q \to \infty$,  one has $\Delta_j \to 0$.
In this limit, $\rho$ and $\rho + \sigma$ represent the classical supertube
radii. The classical un-bubbled solution to which this bubbled solution corresponds
was first  constructed in \refs{\GauntlettQY,\GauntlettWH}.

\goodbreak\midinsert
\vskip .2cm
\centerline{ {\epsfxsize 3.2in\epsfbox{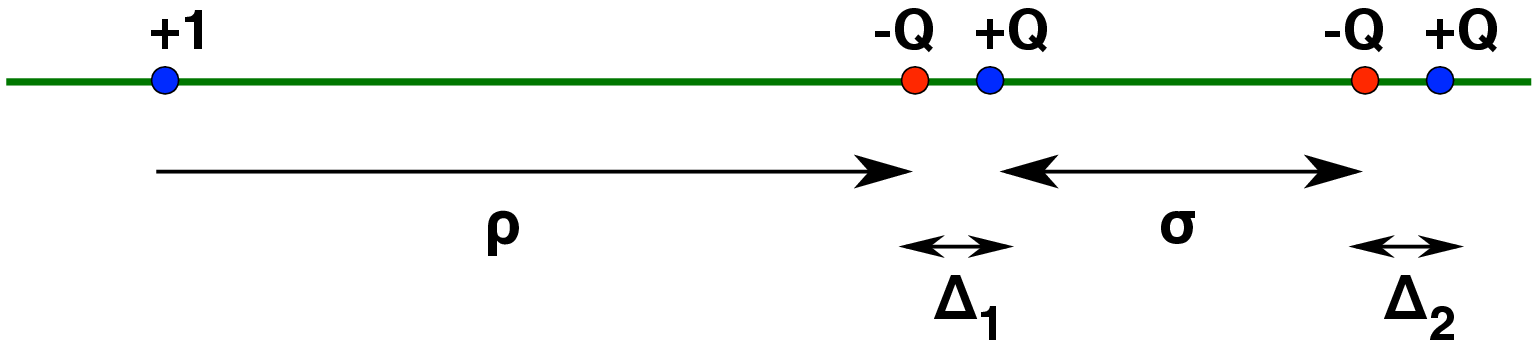}}}
\vskip -.25cm
\leftskip 2pc
\rightskip 2pc\noindent{\ninepoint\sl \baselineskip=8pt
{\bf Fig.~1}:  The layout of GH points for two bubbled supertubes. }
\endinsert

As usual we denote the flux parameters by
\eqn\fluxperiods{ \Pi^{(I)}_{ij}  ~=~   \bigg( {k_j^I \over q_j} ~-~
{k_i^I \over q_i} \bigg) \,,}
but there is a more natural, gauge invariant basis of flux parameters
given by:
\eqn\dfring{\eqalign{d_1^I ~\equiv~ & 2 \,(k_2^I ~+~ k_3^I) \,, \qquad
 f_1^I  ~\equiv~ 2\,  k_1^I ~+~ \big(1+  \coeff{1}{Q} \big)\, k_2^I ~+~
\big (1-  \coeff{1}{Q} \big)\,k_3^I    \,, \cr
d_2^I ~\equiv~ & 2 \,(k_4^I ~+~ k_5^I) \,, \qquad
 f_2^I  ~\equiv~ 2\,  k_1^I ~+~ \big(1+  \coeff{1}{Q} \big)\, k_4^I ~+~
\big (1-  \coeff{1}{Q} \big)\,k_5^I    \,.}}
In the classical supertube limit, where  $\Delta_j \to 0$,  the $d^I$
reduce to the number, $n^I$, of M5 branes around the ring profile.
In the GH metric these supertubes are point-like in the $\IR^3$ base and
run around the $U(1)$ fiber.   The parameters, $f_1^I$ and $f_2^I$, are a little
more physically ambiguous
but we have chosen them to be the gauge-invariant combinations
of flux parameters that are made out of the flux parameters associated
to the two separate rings.

It is also easy to see how to define the $d_a^I$ more generally in terms
of the cohomology.   Recall that the homology cycle, $\Delta_{ij}$, can be  defined by
the $U(1)$ fiber running along any curve between $q_i$ and $q_j$.   The
fluxes  through $\Delta_{23}$ and $\Delta_{45}$ are simply ${Q \over 2} d_1^I$ and
${Q \over 2} d_2^I$ respectively.   We also claim that $\Delta_{23}$ and $\Delta_{45}$
are homologous to the Gaussian surfaces that measure the M5-brane fluxes in the classical,
supertube limit.  To see this, first  recall that for the GH points aligned along the $z$-axis
we may take:
\eqn\Aform{ \vec A  \cdot d \vec y~=~ \sum_{j=1}^5 \,  q_j
{(z -z_j) \over r_j}  \, d \phi \,.}
where $\phi$ is the angle in the  $(x,y)$ plane.  In particular,
if  $V = {1 \over r}$ and $\theta$ denotes the polar angle away from the
$z$-axis then the GH metric reduces to that of  $\IR^4 = \IR^2 \times \IR^2$:
\eqn\uvmetric{ds_4^2 ~=~ (du^2 ~+~ u^2 \, d \theta_1^2) ~+~
(dv^2 ~+~ v^2 \, d \theta_2^2)  \,,}
via the coordinate transformation:
\eqn\uvcoords{\eqalign{u ~=~ \coeff{1}{4} \, r^2\, \cos \coeff{\theta}{2}  \,, \qquad \theta_1 ~=~
\coeff{1}{2} \,(\psi +\phi)\,, \cr
v ~=~ \coeff{1}{4} \, r^2\, \sin\coeff{\theta}{2}  \,, \qquad \theta_2 ~=~
\coeff{1}{2} \,(\psi -\phi) \,.}}
Now observe that if one moves along the $z$-axis then the $U(1)$
fiber direction, $(d\psi +A)$, is equal to $2\, d \theta_1= d\psi + d\phi$  in
the ``long intervals'' from $q_1$ to $q_2$,  from $q_3$ to $q_4$ and
from $q_5$ to infinity.  In the supertube limit, the $U(1)$ fiber, and
hence the supertube, lies in the $(u,\theta_1)$ plane.  The Gaussian
surfaces used to define M5-brane charge can be chosen to so that $\theta_1$ is
fixed and $\theta_2$ is varying.   This means they cannot involve   $\Delta_{12}$ and
$\Delta_{34}$, which necessarily involve the fiber direction with co-tangent,
$(d\psi +A) \sim d\theta_1$,  and so the M5-brane Gaussian surfaces can only be
related to $\Delta_{23}$ and $\Delta_{45}$.

The charges and angular momenta of individual black rings are given by:
\eqn\Qring{Q_I^{(a)} ~=~ C_{IJK} \, d_a^J \, f_a^K\,,}
\eqn\JRring{ j_R^{(a)}  ~\equiv~ \coeff{1}{2} \, C_{IJK}\, \big(f_a^I f_a^J d_a^K ~+~
f_a^I  d_a^J  d_a^K\big)  ~-~ \coeff{1}{24} \,(1-Q^{-2})\, C_{IJK}\,  d_a^I d_a^J d_a^K \,,}
\eqn\JLring{j_L^{(a)} ~\equiv~ \coeff{1}{2} \, C_{IJK}\, \big(d_a^I f_a^J f_a^K ~-~
f_a^I d_a^J d_a^K \big)  ~+~ \Big( {3 \,Q^2 - 4\,Q +1 \over 24\, Q^2} \Big) \,
C_{IJK}\,  d_a^I d_a^J d_a^K\,,}
for $a =1,2$.

For the configuration described above and depicted in Fig.~1 we have:
\eqn\Qcomb{Q_I ~=~ Q_I^{(1)}  ~+~ Q_I^{(2)}  ~+~  C_{IJK} \, d_1^J \, d_2^K\,,}
\eqn\JRcomb{J_R   ~=~  j_R^{(1)}  ~+~  j_R^{(2)}  ~+~  d_1^I \,  Q_I^{(2)} ~+~
d_2^I \,  Q_I^{(1)} ~+~ \coeff{1}{2}\, C_{IJK} \, d_1^I \, d_2^J \, (d_1^K  +
d_2^K) \,,}
\eqn\JLcomb{J_L   ~=~  j_L^{(1)}  ~+~  j_L^{(2)}  ~+~  d_1^I \,  Q_I^{(2)} ~-~
d_2^I \,  Q_I^{(1)} ~+~ \coeff{1}{2}\, C_{IJK} \, d_1^I \, d_2^J \, (d_1^K -
d_2^K) \,,}

It is useful to introduce the flux vectors
\eqn\Ydefn{Y^I    ~\equiv~   \big( f_2^I -  f_1^I  -\coeff{1}{2} \,(d_2^I -  d_1^I ) \big)
 \,,}
and the combination of fluxes:
\eqn\Phatdefn{  \widehat P ~\equiv~
\big(P_{24}  ~-~ P_{25}  ~-~ P_{34}  ~+~ P_{35} \big)~=~
  \coeff{1}{8\, Q^2} \, C_{IJK}\, d_1^I \, d_2^J \, Y^K \,.}
Note that $\widehat P$ measures the total flux running between the
pairs of points that define the two rings.
The interaction part of the left-handed angular momentum can now be written:
\eqn\JLint{\eqalign{J_L^{int}   ~\equiv~ & d_1^I \,  Q_I^{(2)} ~-~
d_2^I \,  Q_I^{(1)} ~+~ \coeff{1}{2}\, C_{IJK} \, d_1^I \, d_2^J \, (d_1^K -
d_2^K)   \cr
~=~ & 8\, Q^2 \, \widehat P ~=~   C_{IJK}\, d_1^I \, d_2^J \, Y^K    \,.}}
From a four-dimensional perspective, the angular momentum $J_L^{int}$ corresponds to the Poynting vector coming from the interaction of the electric fields of one ring with the magnetic fields of the other. We will see in the next sub-section that this controls the merger of the two rings.

\subsec{Classical limits and their entropy}

For a single, classical black ring, the entropy is given by
\eqn\RingEnt{
S ~=~   \pi \sqrt{\cal M} \,}
where
\eqn\cMdefn{\eqalign{
{\cal M} ~=~  &     2\, d^1 d^2  Q_1  Q_2 +2\,d^1 d^3  Q_1  Q_3   +2\,d^2 d^3  Q_2  Q_3
-  (d^1  Q_1)^2 -  (d^2 Q_2)^2  -  (d^3  Q_3)^2  \cr
& ~-~    d^1 d^2 d^3\, \big[4\,  J_L ~+~ 2\, (d^1 Q_1 + d^2 Q_2 + d^3 Q_3)  ~-~
3\,  d^1 d^2 d^3  \big] \,,}}
where $J_L >0$, the $d^I$ are the numbers of M5 branes and the $Q_I$ are the electric
charges  measured from infinity.  One also has the following relation between the angular
momentum, $J_L$, and the classical embedding radius, $R$, measured in $\IR^2$:
\eqn\JLRreln{
J_L  ~=~    (d^1 + d^2 +  d^3)\, R^2 \,.}

If one substitutes the expressions, \QIchg\ and  \JLring, for the charges and  for the angular
momentum of a  single, bubbled ring into \cMdefn, one obtains a simple expression:
\eqn\cMbubbled{
{\cal M} ~=~    \Big( { 4\,Q - 1 \over  Q^2} \Big)   \, (d^1 d^2 d^3)^2  \,.}
Observe that this vanishes as $Q \to \infty$.  This is the ``classical limit'' where the bubbled
ring  collapses back to the standard, classical ring.  Therefore, the classical object corresponding
to this simple, bubbled configuration has $\cM=0$ and is thus a supertube.

For a bubbled ring, the relation, \JLRreln, emerges from the bubble
equations as \BenaVA:
\eqn\JLRrelnbubbled{
J_L  ~=~    4\, (d^1 + d^2 +  d^3)\, \rho \,.}
where $\rho$ is the ring radius measured in the GH base, and the change of variable
\uvcoords\  leads to $\rho = {1 \over 4} R^2$.

If one merges two bubbled supertubes so as to obtain a single bubbled ring, one has an object
with M5 brane charge given by $d^I  =d_1^I + d_2^I$ and with charges and angular momenta
given by  \Qcomb,\JRcomb\ and  \JLcomb.  To obtain the entropy of the corresponding
classical object, one substitutes these expressions into  \cMdefn\ and the result is:
\eqn\cMmerged{\eqalign{
{\cal M} ~=~   &-  \big(\epsilon_{IJK} \, d_1^I d_2^J  Y^K\big)^2  ~-~ 4\, \Big[(d_1^1+d_2^1)
(d_1^2 d_1^3 d_2^1 + d_2^2 d_2^3 d_1^1) \,Y^2  \,Y^3 \cr
&+~  (d_1^2+d_2^2) (d_1^1 d_1^3 d_2^2 + d_2^1 d_2^3 d_1^2) \,Y^1  \,Y^3   + (d_1^3+d_2^3)
(d_1^1 d_1^2 d_2^3 + d_2^1 d_2^2 d_1^3)  \,Y^1 \,Y^2   \Big] \cr
&-~  \coeff{2}{3}\,\big(C_{IJK}\, (d_1^I+d_2^I) (d_1^J+d_2^J)(d_1^K+d_2^K)\big)  \,
\big(C_{ABC}\, d_1^A d_2^B  Y^C\big)  \cr
&+~ \big( \coeff{ 4\,Q - 1 }{36\,Q^2} \big) \big(C_{IJK}\, (d_1^I+d_2^I) (d_1^J+d_2^J)
(d_2^K+d_2^K)\big) \big(C_{ABC}\,  (d_1^A   d_1^B  d_1^C+d_2^A   d_2^B  d_2^C) \big)  \,.}}
Note that if this is generically non-vanishing  as $Q \to \infty$.  However, if $Y^I =0$ then it does
go to zero as $Q \to \infty$.

Putting it somewhat differently,  if $Y^I =0$ then the merged ring has an effective $d^I$ and
$f^I$  given by:
\eqn\fdeff{
d^I  ~=~      d_1^I  ~+~ d_2^I \,, \qquad   f^I  ~=~   f_1^I  ~+~ \coeff{1}{2}\, d_2^I  ~=~
 f_2^I  ~+~ \coeff{1}{2}\, d_1^I   \,.}
The fact that one adds the $d_a^I$  follows from the considerations in Section 3.1
and the formula for $f^I$ is obtained from \Qcomb\ and \Qring, using $Y^I=0$.
Now observe that for large $Q$, one has
\eqn\jlsimp{
j_L^{(a)} ~\approx~ \coeff{1}{2} \, C_{IJK}\,  d_a^I \big(f_a^J - \coeff{1}{2}\, d_a^J\big)
 \,  \big(f_a^K - \coeff{1}{2}\, d_a^K\big) ~=~ \coeff{1}{2} \, C_{IJK}\,  d_a^I
 \big(f^J - \coeff{1}{2}\, d^J\big)  \,  \big(f^K - \coeff{1}{2}\, d^K\big) \,.}
It then follows that when $Y^I=0$ the angular momentum, $J_L$, for the merged
ring is given by:
\eqn\jlsimp{\eqalign{
J_L   ~=~&   j_L^{(1)}  ~+~  j_L^{(2)}  ~+~    C_{IJK}\, d_1^I \, d_2^J \, Y^K
~=~   j_L^{(1)}  ~+~  j_L^{(2)}
\cr  ~\approx~  & \coeff{1}{2} \, C_{IJK}\,  d^I
 \big(f^J - \coeff{1}{2}\, d^J\big)  \,  \big(f^K - \coeff{1}{2}\, d^K\big) \,,}}
which is the angular momentum, $J_L$, for a bubbled ring or supertube
of charges $d^I$ and $Q_I$.  In other words the merged configuration
still has a maximal value of $J_L$ and the corresponding classical
object still has vanishing horizon area.  If $Y^I \ne 0$ then the final angular
momentum will  generically  be less  than this maximal value\foot{This is not
obvious from \cMmerged\   because one must also require the absence of
CTC's in the solution.  This comment is therefore based primarily on the essential
physics of  mergers as well as experience with a number of examples. }.

\subsec{The bubble equations}

For the configuration depicted in Fig.~1, there are four independent bubble equations
\BubbleEqns.
If one adds the equations for $i=2,3$ and $i=4,5$ then eliminates
terms with denominators $r_{23} = \Delta_1$ and $r_{45} = \Delta_2$, one obtains:
\eqn\twobubbs{  \eqalign{Q\,\Big( {P_{12} \over \rho} - {P_{13} \over \rho + \Delta_1}
\Big)  ~+~  Q^2\,\Lambda  ~=~ & - \coeff{1}{2}\, \sum_{I=1}^3 d_1^I \cr
Q\,\Big( {P_{14} \over \rho+\sigma + \Delta_1}   - {P_{15} \over \rho + \sigma +
\Delta_1+ \Delta_2}   \Big)  ~-~  Q^2\,\Lambda ~=~ &- \coeff{1}{2}\, \sum_{I=1}^3 d_2^I \,.}}
where
\eqn\Lambdadefn{
\Lambda~\equiv~ {P_{24} \over \sigma + \Delta_1 } - {P_{34} \over \sigma} -
{P_{25} \over \sigma + \Delta_1+ \Delta_2}  + {P_{35} \over \sigma +  \Delta_2}\,.}

For two bubbled rings with $\Delta_j \ll \rho, \sigma$, assuming that all other terms in
the multipole expansion are sub-leading, these equations reduce to:
\eqn\tworings{  \eqalign{  {Q\,(P_{12} -  P_{13}) \over \rho}   ~+~
{Q^2\,\widehat P \over \sigma}  ~=~ &-  \coeff{1}{2}\, \sum_{I=1}^3 d_1^I \cr
 {Q\,(P_{14} -  P_{15}) \over \rho+ \sigma}   ~-~
{Q^2\,\widehat P \over \sigma}    ~=~ &- \coeff{1}{2}\, \sum_{I=1}^3 d_2^I \,.}}
For two generic bubbled rings to merge one must have $\sigma \to 0$
and so the merger condition is $\widehat P \to 0$, but with $ \sigma^{-1} \widehat P $
remaining finite.  Note that this means  that the interaction part of the left-handed angular
momentum, $J_L^{int}$,  must vanish.  More generally, for a  solution in which
 $\Delta_j$ and $\sigma$ get small simultaneously, one must have
$\widehat P \to 0$, but with $\Lambda$  remaining finite.  Either way, the location,
$\rho_0$, of the merged object is given by:
\eqn\mergerho{\eqalign{\rho_0 & ~=~    -2\, Q\, \bigg[\sum_{I=1}^3(d_1^I  + d_2^I )
 \bigg]^{-1}\, \big(P_{12} -  P_{13} + P_{14} -  P_{15}\big) \cr
&  ~=~  \coeff{1}{4} \, \bigg[\sum_{I=1}^3(d_1^I  + d_2^I ) \bigg]^{-1}\,  \Big(
j_L^{(1)} ~+~j_L^{(2)} ~+~   \coeff{1}{6\,Q} \, C_{IJK}\,  \big(d_1^I d_1^J d_1^K +
d_2^I d_2^J d_2^K\big) \Big)\,,}}
which is equivalent to \JLRrelnbubbled\ for the combined object.

Conversely, if one has $\widehat P \to 0$ one can obtain the merger solution,
and a second solution in which $\sigma$ remains finite.  One then has:
\eqn\mergerhosig{\eqalign{\rho  & ~=~    -2\, Q\, \bigg[\sum_{I=1}^3 d_1^I
 \bigg]^{-1}\, \big(P_{12} -  P_{13} \big) ~=~   \coeff{1}{4} \, \bigg[\sum_{I=1}^3  d_1^I
  \bigg]^{-1} \Big( j_L^{(1)}   ~+~   \coeff{1}{6\,Q} \, C_{IJK}\, d_1^I d_1^J d_1^K   \Big) \,,  \cr
\rho + \sigma & ~=~    -2\, Q\, \bigg[\sum_{I=1}^3 d_2^I
 \bigg]^{-1}\, \big(P_{14} -  P_{15} \big) ~=~   \coeff{1}{4} \, \bigg[\sum_{I=1}^3  d_2^I
  \bigg]^{-1} \Big( j_L^{(2)}   ~+~   \coeff{1}{6\,Q} \, C_{IJK}\, d_2^I d_2^J d_2^K   \Big) \,.}}
Note that these are essentially the radii given by \JLRrelnbubbled\ for each
of the two rings separately.  Indeed, the  limit $\widehat P \to 0$ corresponds
to the vanishing of the ``interaction part'' of $J_L$.

In the foregoing merger analysis we have assumed that $\Delta_j \ll \rho, \sigma$ and
we have dropped terms from the multipole expansions of the denominators
in \twobubbs.   This means that the fluxes through the rings, $\Pi_{23}^{(I)}$ and
 $\Pi_{45}^{(I)}$,  must be much less than the other fluxes, and so $ Q^{-1} d^I_a $
 must be small compared to $f^I_a$ and $d^I_a$.  Thus we should  consistently
 drop terms that are sub-leading in $ Q^{-1} d^I_a $, like the last terms in
\mergerho\ and \mergerhosig.  Such terms also appear as corrections coming from
the multipole expansions \BenaKB.

\newsec{Scaling solutions}

A scaling solution is most simply defined to be a bubble configuration where
there is a subset, $\cS$, of the GH points that are uniformly approaching one
another as some control parameter or, perhaps a modulus, is adjusted
to a critical value.  That is, one has
\eqn\scaling{  r_{ij} ~\to~ \lambda\,  r_{ij}\,, \qquad i,j \in \cS\,,}
with $\lambda \to 0$.
Physically, if the total charge of the GH points in $\cS$ is non-zero,
this means that these GH points are descending into an arbitrarily deep
black-hole-like throat and that the red-shifts of excitations localized around
these points are going to infinity.  Such black-hole microstates are called
``Deep Microstates'' and were  discovered in  \BenaKB\ via the
study of mergers of black holes and black rings \BenaZY.  In particular, scaling solutions
were shown  to be associated to microstates of black holes of non-zero horizon area
and it was also argued (using the dual CFT) that these deep microstates belong to the same sector as
 the {\it typical} microstates of the black hole.

Here we examine the corresponding story when the total GH charge in $\cS$ is
zero.  One then expects to obtain scaling solutions corresponding to
deep microstates of black rings with non-zero horizon area.  This is, indeed
what we find.

So far, the study of scaling solutions has largely focussed
on $U(1) \times U(1)$ invariant configurations.  This is largely because
such solutions are intrinsically simpler and, for fixed fluxes, there are finitely many,
discrete solutions that are $U(1) \times U(1)$ invariant.  Moreover, in scaling solutions
with such a symmetry, the depth of the throat is controlled by the choice of the
quantized fluxes on bubbles \BenaKB, like $\widehat P$ in Section 3.  Thus
the size of the flux quanta provides a cut-off for the maximum depth of the throat.
However, one can easily break  the $U(1) \times U(1)$ symmetry to $U(1)$
by letting the GH points move to arbitrary points, $ \vec r_i$, in $\IR^3$ and then a
subset of the angles between the vectors, $\vec r_{ij} = \vec r_j - \vec r_i$,
become continuous moduli of the solutions.  If the initial fluxes lie in the right domain
then one can find scaling solutions at special points, or on special surfaces of
the moduli space.   Thus, we can make the black-hole, or black-ring throat arbitrarily
deep by tuning the moduli.  In practice, we find (numerically) that this tuning
has to be extremely sharp and that the throat depth varies by many orders of magnitude
for tiny (micro-radian) variations of the angles.

\subsec{Axi-symmetric scaling solutions}

Since the $U(1) \times U(1)$ invariant solutions are discrete
for given fluxes, we can achieve an axi-symmetric scaling solution only
by delicate adjustment of the fluxes.  In particular, the easiest way to
find such a scaling solution is via the merger of two bubbled supertubes
as described in Section 3.  Specifically, we need $J_L^{int} \to 0$, which
means that the fluxes need to be adjusted so that $\widehat P \to 0$.
To get a deep microstate, the expression for the classical horizon area, \cMmerged,
shows that we must do this in such a  manner that $Y^I$ remains finite and so the
merger condition, \JLint,  implies that we must tune the $f^I$ or the
$d_a^I$  so that $Y^I$ is finite  but orthogonal to $C_{IJK}d_1^J d_2^K$.

It is relatively easy to see that there are scaling solutions that arise through
mergers of bubbled supertubes.  We consider, once again, the configuration
in Fig.~1 where $\Delta_j, \sigma \to 0$.  There are four independent bubble equations,
\BubbleEqns,  to satisfy. (Remember that the sum of the five bubble equations
is trivial.) The first equation is precisely the same as the sum of the two equations in
\twobubbs\ and it determines the position, $\rho_0$, of the merger, as in \mergerho.
For the scaled merging of bubbled supertubes we expect that  $\Delta_j \ll \sigma $
during the merger and so the three remaining, independent bubble equations
have the form:
\eqn\scaletworings{\eqalign{ - {P_{25} \over \sigma +\Delta_1+\Delta_2} ~-~
 {P_{23} \over \Delta_1 }  ~+~  {P_{24} \over \sigma +\Delta_1 } ~\approx~
 {(P_{24}- P_{25} ) \over \sigma  } ~-~
 {P_{23} \over \Delta_1 }     & ~=~   C_1 \,,  \cr
{P_{23} \over \Delta_1 } ~+~  {P_{35} \over \sigma  +\Delta_2} ~-~
 {P_{34} \over \sigma}  ~\approx~  {(P_{35}- P_{34} ) \over \sigma  } ~+~
 {P_{23} \over \Delta_1 } & ~=~   C_2  \,,  \cr
- {P_{24} \over \sigma +\Delta_1 }  ~+~ {P_{34} \over \sigma }  ~-~
{P_{45} \over  \Delta_2} ~\approx~  {(P_{34}- P_{24} ) \over \sigma  } ~-~
 {P_{45} \over \Delta_2 } & ~=~   C_3 \,,  }}
where the $C_j$ contain terms involving the fluxes,  $Q$ and $\rho_0$.
The important point about \scaletworings\ is that we have explicitly
shown the terms that grow large as $\Delta_j, \sigma \to 0$.
For the scaling solution we need:
\eqn\scalestwo{ \sigma ~=~ \lambda\, \sigma^{(0)}\,, \qquad
\Delta_a ~=~ \lambda\, \Delta_a^{(0)}\,, \qquad \lambda\to 0\,.}
This means that, to leading order as $\lambda \to 0$:
\eqn\ratiostwo{  {\sigma^{(0)} \over \Delta_1^{(0)} }  ~\approx~   {(P_{24}- P_{25} ) \over
P_{23}  } ~\approx~   {(P_{34}- P_{35} ) \over
P_{23}  }   \,, \qquad  {\sigma^{(0)} \over \Delta_2^{(0)} }  ~ \approx ~   {(P_{34}- P_{24} ) \over
P_{45}  } \,.}
In particular there is no conflict between the first and second equations in
\scaletworings\ because $\widehat P \to 0$.
The foregoing solution becomes more and more accurate for smaller values of
$\lambda$ and given this solution one can then easily find the solution for
finite, larger values of $\lambda$ using perturbation theory.

This is not necessarily the only scaling solution with the configuration shown
in Fig.~1.  Indeed, numerical solutions show that there is often another one
in which the $\Delta_j$ and $\sigma$ are of approximately the same order.
However in all the examples of this second solution that we have found, there are
large regions of CTC's.
On the other hand, the scaling solutions that we have found based upon mergers
of bubbled supertubes appear to be free of CTC's.
  We will discuss this more in Section 4.2.

\subsec{Numerical results for axi-symmetric scaling solutions}

In order to see the scaling solutions explicitly, and verify that there are no CTC's
we constructed several numerical examples and we now discuss a representative
case.

It is useful to define:
\eqn\XIdefn{ X_a^I ~\equiv~  f_a^I ~-~ \coeff{1}{2} \, d_a^I\,.}
We then take $Q=105$ and:
\eqn\fluconfnum{\eqalign{ d_1^I ~=~& (\, 50\,,\, 60\,,\, 40 \,) \,, \quad X_1^I ~=~
(\, 110\,,\, 560\,,\, 50 \,) \,, \cr d_2^I ~=~& (\, 80\,,\, 50\,,\, 45 \,) \,, \quad X_2^I ~=~
(\, x \,,\, 270\,,\, 280 \,) \,,}}
where $x$ is varied from about $64$  up to its merger value of $x \approx 90.3$.
The results are shown in Fig.~2.   As is evident from the graph, there are three
sets of solutions to the bubble equations.   Branch (i)   exists for all values of $J_L^{int}$
and has all four GH points  in a very close cluster that scales as  $J_L^{int} \to 0$.
This is appears as a very steep line at the center of Fig. 2.  Branches (ii) and (iii)
only appear at a bifurcation point when one has  $|J_L^{int}|  \lesssim 43,500$ or
$x \gtrsim 71.7$,   and represent solutions in which the four GH points separate
into two sets of very  close pairs.  On branch (ii) the two pairs move apart as
$J_L^{int} \to 0$ and the
locations of the two bubbled rings is given by \mergerhosig.   Branch (iii)
is the scaling merger solution at $\rho_0$ given by  \mergerho\  and is described
by \scalestwo\ and \ratiostwo.

We have done extensive numerical searches for CTC's in all these solutions
and we find that branch (i) is completely unphysical, with large regions of CTCs,
but that branches (ii) and (iii) are physical and have no CTC's.

\goodbreak\midinsert
\vskip .3cm
\centerline{ {\epsfxsize 3.5in\epsfbox{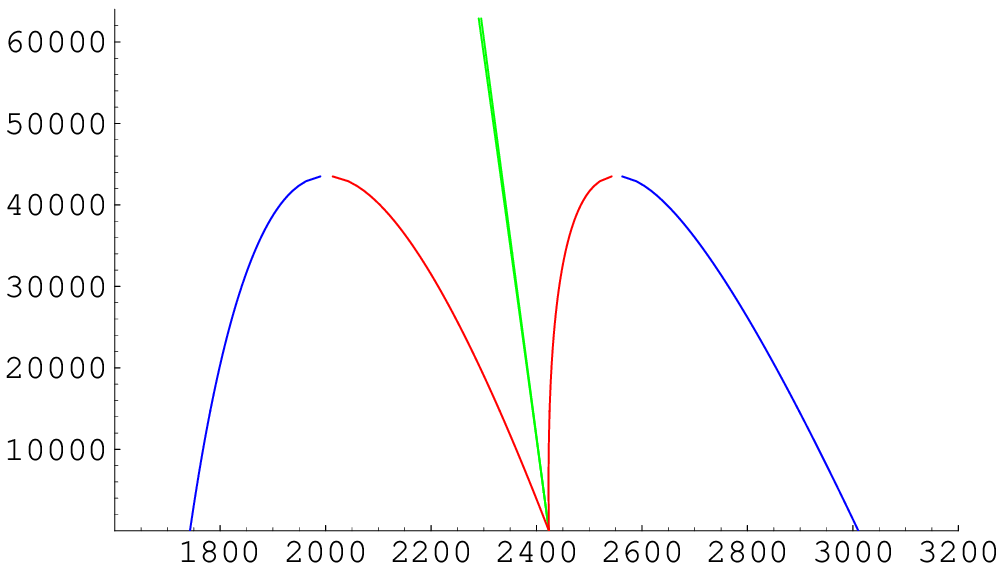}}}
\vskip 0.2cm
\leftskip 2pc
\rightskip 2pc\noindent{\ninepoint\sl \baselineskip=8pt
{\bf Fig.~2}:  Solutions of the bubble equations for the configuration
shown in Fig.~1.  The plot shows the ring positions, $\rho$ and $\sigma$,
along the horizontal axis with $|J_L^{int}|$ plotted on the vertical axis.
The separations, $\Delta_j$, are too small to resolve.
There are three branches: (i) The single, nearly  vertical line in the center (in green)
for which all four GH points remain extremely close together;
(ii)  The two outermost curves (in blue)
where the two rings become progressive more widely spaced as
$J_L^{int} \to 0$; (iii)  The two curves (in red)  that meet branch (ii) and
show the scaling merger in which the two rings meet at    $J_L^{int} =  0$.}
\endinsert

Finally, one can use \cMdefn\  or \cMmerged\  to compute the horizon area,
$\cM$,  of a classical black ring with the same charges and angular momenta as
the merged configuration.   The absolute number does not immediately convey
useful information.  On the other hand, we can compare this to the ``maximal
horizon area,'' $\cM_0$, of a black ring with the same values of $Q_I$ and
$d^I$, but with $J_L =0$.  For the configuration in \fluconfnum\ we find:
\eqn\nonextredef{{\cal{R}}~\equiv~ {{\cal{M}} \over {\cal{M}}_0} ~\approx~ 0.14\,, }
Thus the the result of this merger of two bubbled supertubes is a microstate
of a black ring with a non-vanishing horizon area.

We have studied several other such mergers with different values of
flux parameters and found a number of solutions that are free of CTC's
and have even higher values of $\cR$.  Indeed, one can arrange a very
high value of $\cR$ if one takes the outer ring to rotate in the opposite
direction to the inner ring.  One can achieve this in the foregoing
example, \fluconfnum, by taking, for example,
\eqn\newXtwo{   X_2^I ~=~  (\, -300 \,,\, 270\,,\, 531.27 \,) \,,}
while leaving all the other parameters unchanged.  This configuration
is very close to the merger point and has $\cR \approx 0.638$.
As one would expect, one generates more entropy by merging
states whose angular momenta are opposed to one another.

\newsec{Abysses and closed quivers.}

Thus far we have primarily focussed on  $U(1) \times U(1)$ invariant
 scaling solutions.   It is relatively easy to modify the analysis above to obtain
 scaling solutions in which the five GH points no longer lie on an axis.  It is, however,
 even simpler to find scaling solutions based upon four GH points, and this
 is what we will focus on here.

Consider four charges laid out as in Fig.~3 with:
\eqn\fourchgs{q_1 ~=~  +1\,, \qquad q_2 ~=~  2\, Q \,, \qquad q_3 ~=~  -  Q\,, \qquad
q_4 ~=~  -Q  \,.}
The general bubble equations take the form:
\eqn\tribubb{\eqalign{{2\, Q\, P_{12} \over r_{12}}   ~-~  {Q\, P_{13} \over r_{13}}
  ~-~  {Q\, P_{14} \over r_{14}}  ~=~
- \sum_{I=1}^3 \, \tilde k_1^I & ~=~   C_1 \,,  \cr
- {2\, Q\, P_{12} \over r_{12}} ~-~
{2\, Q^2\, P_{23} \over r_{23}}   ~-~  {2\, Q^2\, P_{24} \over r_{24}}  ~=~
- \sum_{I=1}^3 \, \tilde k_2^I & ~=~    C_2 \,,  \cr
{Q\, P_{13} \over r_{13}}    ~+~
{2\, Q^2 \, P_{23} \over r_{23}}  ~+~  {Q^2\, P_{34} \over r_{34}}  ~=~
- \sum_{I=1}^3 \, \tilde k_3^I & ~=~    C_3  \,,  \cr
 {Q\, P_{14} \over r_{14}}  ~+~  {2\, Q^2\,P_{24} \over r_{24}}   ~-~
{ Q^2 \, P_{34} \over r_{34}}  ~=~
- \sum_{I=1}^3 \, \tilde k_4^I & ~=~   C_4 \,,  }}
%

\goodbreak\midinsert
\vskip .2cm
\centerline{ {\epsfxsize 3.6in\epsfbox{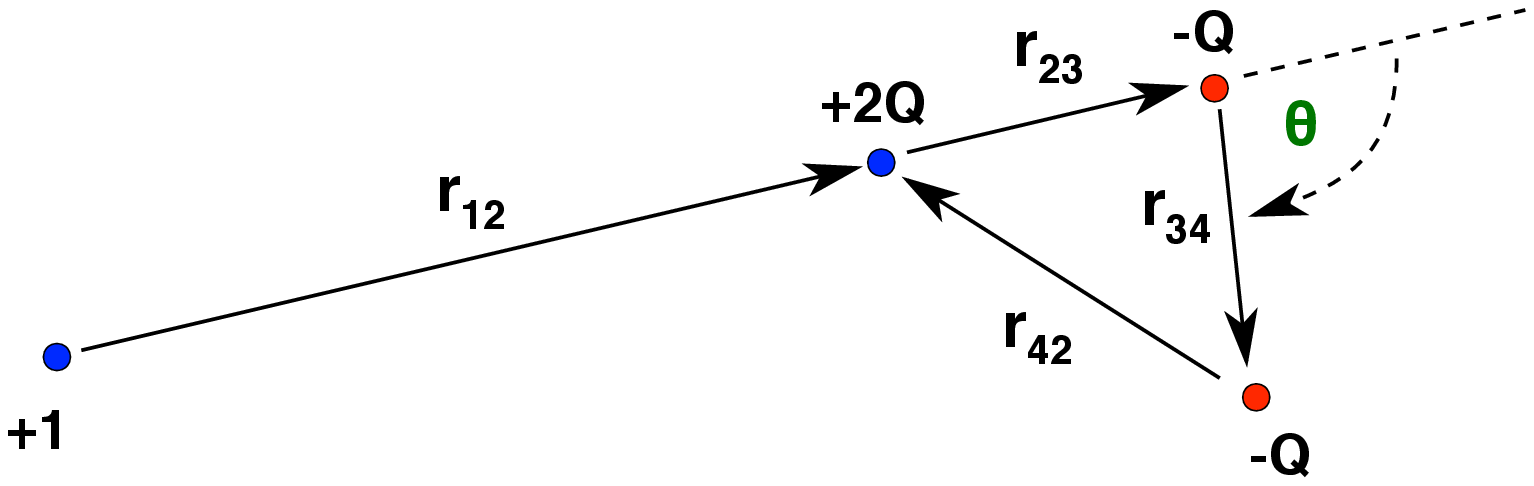}}}
\vskip -.25cm
\leftskip 2pc
\rightskip 2pc\noindent{\ninepoint\sl \baselineskip=8pt
{\bf Fig.~3}:  The layout of GH points for a triangular scaling solution. }
\endinsert

To obtain a scaling solution whose classical limit is a ring, the triangle defined
by points $2,3$ and $4$ should collapse so that:
\eqn\limitrij{r_{1j} \approx \rho \,, \qquad r_{ij} ~\ll~ \rho \,, \qquad  i,j = 2,3,4\,.}
Indeed, for a scaling solution in which $r_{ij} \to 0$ for $ i,j = 2,3,4$, the bubble
equations \tribubb\ require that:
\eqn\limitrij{r_{ij} ~\to~ (-1)^{i+j+1} \lambda\,  q_i \, q_j \, P_{ij}     \,, \qquad
2 \le i < j \le 4\,,}
with $\lambda \to 0$.  In other words, the fluxes define the lengths of the
sides and hence the angles in the triangles.  One then has:
\eqn\JLijlim{\vec J_{L\, ij} ~=~ -  8\, (-1)^{i+j+1}  \lambda \,  \vec r_{ij}
 \,, \qquad   2 \le i < j \le 4\,,}
where $\vec r_{ij} \equiv (\vec y^{(i)} - \vec y^{(j)})$.  It follows that
\eqn\JLintlim{ \vec J_{L}^{int}~=~ \sum_{i,j=2}^4 \, \vec J_{L\, ij} ~\to~
- 16\, \lambda \, (\vec r_{23} +\vec r_{34} -\vec r_{24} )  ~\equiv~0\,,}
because these vectors define the sides of the triangle.  The last bubble
equation then yields:
\eqn\newringrad{\rho ~\approx~  2\, Q\, \bigg[\sum_{I=1}^3  d^I \bigg]^{-1} \,
 \big( P_{12}  ~-~ 2\,  P_{13} ~+~ P_{14} )\,,}
where
\eqn\ddefntri{d^I ~\equiv~  2\, \big(k_2^I ~+~k_3^I ~+~k_4^I \big) \,.}
Moreover, one also finds that the combination of fluxes in \newringrad\ is exactly
the non-vanishing part of $J_L$ and so one, once again, recovers \JLRrelnbubbled.

Thus we find scaling solutions for generic values of fluxes:  The only
constraint is that the $|q_i   q_j P_{ij}| $ must satisfy the triangle inequalities.
Now recall that only three of the equations  in \tribubb\ are independent.  One of
them fixes $\rho$ and the others fix the lengths of two sides of the triangle in
terms of the length of the third side.  Thus we may view the angle, $\theta$, in
Fig.~3 as a modulus of the solution.   The scaling solution then appears
when the angle, $\theta$, is tuned so that the triangle has the shape
determined by the fluxes as in \limitrij.
The new feature of this class of solutions is that we are no longer fine-tuning a
quantized  flux parameter in order to approach the scaling limit.   For these triangular
scaling solutions, one can pick the quantized fluxes and then the scaling solution
appears as a {\it modulus} is tuned to a critical value.

Obviously, not all of these triangular scaling solutions will be free of
CTC's and this will put further constraints on the flux parameters.  However, we
have found a number of numerical examples that exhibit scaling at the critical
value of $\theta$ and reveal no CTC's under careful numerical scrutiny
of the solution.   One such example has the following parameters:
\eqn\trichargefluxpa{\eqalign{q_i ~=~& (1\,, 210\,,\, -105\,,\,\,  -105)\,, \quad
k_i^1 ~=~ (0\,,\,  525\,,\, 1200\,,\, 2210) \,, \cr k_i^2 ~=~&
(0\,,\, -20000\,,\, 16000\,,\, 7887)\,, \quad k_i^3 ~=~ (0\,,\,   6400\,,\, 1613\,,\, 7900)\,,}}
where, $i=1,\dots,4$.   Define $\Gamma_{ij} ~=~ q_i q_j P_{ij}$ then we have:
\eqn\triflux{\Gamma_{23} ~=~  8.0446 \times 10^8\,, \quad \Gamma_{34} ~=~
4.9063 \times 10^8\,, \quad \Gamma_{24} ~=~ -1.1046 \times 10^9\,.}
Note that the magnitudes of these fluxes all satisfy the triangle inequalities
\eqn\triangineq{|\Gamma_{ij} | ~\le ~  |\Gamma_{ik} | ~+~ |\Gamma_{kj} | \,.}
By solving the bubble equations numerically, we find the overall size of the
ring blob depends on the shape of the triangle formed by the three charges.
In Table 1 we show how the size of the triangle changes as we vary the angle,
$\theta$.

The total dipole charges of this merger solution are given by:
\eqn\diptotal{d^I ~\equiv~ 2\,\sum_{j=1}^3\, k_j^I \,,}
while the electric charges and angular momenta can be obtained from
\QIchg, \Jright\ and \Jleft.  From these one can obtain the horizon area
ratio, ${\cal R}$, in \nonextredef\  for the corresponding classical black-ring
solutions and here we find  $\cR \approx 0.103$.  Thus this scaling solution
represents a microstate of a true black ring.

\goodbreak
{\hskip -1.25cm \vbox{\ninepoint{
$$
\vbox{\offinterlineskip\tabskip=0pt \halign{\strut\vrule# &\vrule\hfil ~~# \hfil &\vrule # &\hfil ~~#~~ \hfil &\vrule # &\hfil ~~#~~ \hfil &
\vrule # &\hfil ~~#~~ \hfil &\vrule # &\hfil ~~#~~ \hfil & \vrule # &\hfil ~~#~~ \hfil & \vrule # &\hfil ~~#~~ \hfil& \vrule # \cr
\noalign{\hrule} & ~$\pi\,-\,\theta $&&$r_{12}$&&$r_{23}$&&$r_{34}$&&$r_{24}$&&${r_{34}/r_{23}}$ &&${r_{24}/r_{23}}$ & \cr
\noalign{\hrule height1.5pt}
& 0 && 580.889&& 28.601 && 19.150 && 47.751 && .66954&& 1.6695& \cr
\noalign{\hrule}
& .2 && 532.623&& 27.820 && 18.577 && 46.175 && .66776&& 1.6598& \cr
\noalign{\hrule}
& .4 && 537.742&& 25.439 && 16.851 && 41.482 && .66238&& 1.6306& \cr
\noalign{\hrule}
& .6 && 546.005&& 21.341 && 13.943 && 33.779 && .65333&& 1.5828& \cr
\noalign{\hrule}
& .8 && 557.025&& 15.323 && 9.8136 && 23.251 && .64046&& 1.5174& \cr
\noalign{\hrule}
& 1 && 570.302&& 7.0865 && 4.4196 && 10.178 && .62366&& 1.4363& \cr
\noalign{\hrule}
& 1.1 && 577.612&& 2.0172 && 1.2381 && 2.8049 && .61375&& 1.3905& \cr
\noalign{\hrule}
& 1.13 && 579.878&& .36089 && .22035 && .49664 && .61058&& 1.3762& \cr
\noalign{\hrule}
& 1.136 && 580.335&& .021823 && .013311 && .029969 && .60993&& 1.3732& \cr
\noalign{\hrule}
& 1.13635 && 580.361&& 1.963 $\times 10^{-3}$&& 1.197 $\times 10^{-3}$&& 2.695 $\times 10^{-3}$&& .60990&& 1.3731& \cr
\noalign{\hrule}
& 1.136383 && 580.364&& 9.008 $\times 10^{-5}$&& 5.494 $\times 10^{-5}$&& 1.237 $\times 10^{-5}$&& .60989&& 1.3731& \cr
\noalign{\hrule}
& 1.136384586 && 580.364&& 6.289 $\times 10^{-8}$&& 3.836$\times 10^{-8}$&&8.635 $\times 10^{-8}$&& .60989&& 1.3731& \cr
\noalign{\hrule}
& 1.1363845871 && 580.364&& 4.573 $\times 10^{-10}$&& 2.789 $\times 10^{-10}$&& 6.279 $\times 10^{-10}$&& .60989&& 1.3731& \cr
\noalign{\hrule}
&1.136384587108&& 580.364&& 3.207 $\times 10^{-12}$&& 1.956 $\times 10^{-12}$&& 4.403 $\times 10^{-12}$&& .60989&& 1.3731& \cr
\noalign{\hrule} }\hrule}$$
\vbox{ \vskip+7pt \leftskip 2pc \rightskip 2pc\noindent{ \ninepoint\sl \baselineskip=8pt
{\bf Table 1}: This table shows the distances,  $r_{ij}$,  between point $i$ and point $j$
in the triangle solution as a function of the modulus, $\theta$. (See Fig. 3.)
The angle, $\theta$, is the angle between $\vec r_{23}$ and $\vec r_{34}$  and it is
varied to produce the merger at  $\theta = \theta_c$ (given below),  while all the other
parameters are kept fixed.    One should also note that the ratios of the
distances at merger are precisely the ratios of the fluxes:   $|\Gamma_{34}/\Gamma_{23}|
\approx 0.609893$ and $|\Gamma_{24}/\Gamma_{23}| \approx 1.37308$.
}}} \vskip7pt}}

As the angle $\theta$ is changed towards a critical angle, given by
\eqn\thcrit{\pi - \theta_{c}  \approx 1.13638458710805705...}
the distances inside the ring blob as measured on the base $(r_{23},r_{34},r_{24})$ shrink as well,
such that
\eqn\scalingth{r_{ij} \sim \theta - \theta_c}
We have checked at great length that this solution is free of closed timelike curves and
has a global time function. Some details of the investigation are presented in the Appendix.

As explained in \refs{\BenaKB,\BenaKG}, during the scaling \scalingth\ the physical
distances between the points that form the cap remains the same, while the throat
becomes longer and longer\foot{We have also checked this explicitly for the
solution presented in Table 1.}. The structure of the cap remains self-similar, the curvature
is bounded above by a cap-dependent value that is parametrically smaller than the
Planck size. Hence, supergravity is a valid  description of the scaling solution for
any throat depth! Interestingly-enough, as $\theta \rightarrow \theta_c$, the length of
the throat diverges, and the solution becomes an ``abyss'' that increasingly resembles
the naive black hole solution.

From a four-dimensional perspective, the solutions we consider here correspond to
multi-centered configurations of D6 and $\overline{\rm D6}$ branes. The fact that four-dimensional
multi-center solutions can collapse has been known for quite a while
\refs{\DenefNB\DenefRU{--}\BatesVX} (see \DenefVG\ for recent progress) and has been associated
to the existence of ``closed quivers'' in the gauge theory describing these configurations.
The discussion of fluxes and angular momenta presented above can also be
obtained from the more general analysis of multi-black hole four-dimensional solutions,
upon restricting to ${\rm D6-\overline{\rm D6}}$ configurations.

It is also possible to argue from a four-dimensional perspective that
even if the points of the quiver appear to collapse, in fact the
distance between these points remains fixed\foot{We thank Eric Gimon
for pointing out this argument to us.}. The four-dimensional
metric is
\eqn\fourdmetric{ds_{4D}^2 =  - {\cal Q}^{-1/2} (dt + \omega)^2 + {\cal Q}^{1/2} (ds_{\IR^3}^2 )}
where
\eqn\Qdef{{\cal Q} ~\equiv~ Z_1\,Z_2 \, Z_3 \, V ~-~
\mu^2 \, V^2 ~\ge~ 0 \,.}
In a scaling solution where the distances between the centers in the
flat $\IR^3$ metric scales like $\Lambda$, the value of the function
${\cal Q}$ in the region of the centers scales like
$1/\Lambda^2$, when the total charge of the scaling centers is that
of a black hole of non-zero entropy. Hence, the physical distance between the
scaling centers remains constant. This four-dimensional analysis also implies
that only centers whose total charge corresponds to a black hole or a
black ring of finite horizon area can form a deep (abyssal) microstate.

Of course, from a four-dimensional perspective all the GH centers are
naked singularities, and one could object that the distances between
these centers are ill-defined. However, the four-dimensional results
are useful because they complement those obtained from the full five-dimensional solution: the
physical distance between the centers remains fixed throughout the
scaling, and the apparent collapse of the centers manifests itself as
the appearance of a throat.

\newsec{The physics of deep microstates and abysses}

We have found the first examples of smooth microstate geometries that
have the same charges, dipole moments and angular momenta as black rings
with a macroscopically large horizon area.  These solutions are identical to black-ring
solutions, both in the asymptotic region, and in the near-horizon region, but
instead of having the infinite throat of classical BPS black rings, they have a
very deep throat that ends in a smooth cap. All the charges of the solution come
from fluxes threading topologically non-trivial cycles at the bottom of the throat.

If we impose $U(1) \times U(1)$ symmetry on the solutions then
the depth of the throat is naturally limited by the size of the flux quanta
and, as in \BenaKB,  we expect the red-shift of low-energy excitations
near the bottom of the throat to yield an energy that matches the mass gap
of the dual CFT.

We have also found that solutions that do not have a $U(1) \times U(1)$ isometry
can have a throat whose length depends both on the fluxes, and on geometric
moduli of the base space. Most particularly, we have obtained {\it abyssal} solutions
in which the depth of the throat can be made arbitrarily large by tuning certain angles
on the base space!   In these scaling solutions, the size of all the cycles remains
finite as the length of the throat becomes larger and larger, and  hence
the solutions can be described using supergravity for arbitrarily lengths of the throat.
While we have only constructed abyssal solutions corresponding to black rings,
it is pretty clear that black hole microstates with this feature could also be constructed
this way.  From a four-dimensional perspective, these solutions correspond to
D6--$\overline{\rm D6}$ solutions that have closed quivers, and hence the
branes appear to get arbitrarily close to each other \refs{\DenefNB\DenefRU\BatesVX{--}\DenefVG}.
Nevertheless, that perspective is misleading: when considering the full
five-dimensional solution, the physical distances between the GH points
corresponding to the D6 branes remains {\it finite} throughout the scaling.  Moreover,
unlike their four-dimensional counterparts, the solutions that we consider are smooth.

The fact that one can construct smooth horizonless solutions that have arbitrarily long
throats poses interesting questions for the interpretation of microstate geometries
from the point of view of the $AdS$/CFT correspondence. Since these geometries are dual
(up to $1/N$ corrections) to states of the boundary CFT, it appears naively that
these states will have an arbitrarily small mass-gap, as well as a whole tower
of excitations that can be made arbitrarily light, contradicting expectations for a
quantum theory in a box.  Moreover, since the geometries we construct are
supersymmetric and have very large cycles, and hence very low curvatures, one can
imagine perturbing them slightly by adding a suitably small box of gas with some entropy,
and doing this without significantly disturbing the
geometries.  If one then dials the length of the throat to become arbitrarily large one
will obtain a system that has the entropy of the gas, but has an energy arbitrarily
close to the BPS bound.  In the following sub-sections we will refine these puzzles and
discuss some possible resolutions.

\subsec{The spectrum and mass gaps in $AdS$/CFT:  The puzzle}

The best-studied theory that is holographically dual to the geometries we consider
is the D1-D5 CFT.   At strong coupling this CFT is dual to string theory on
$AdS_3 \times S^3 \times T^4$. Even though our geometries are constructed in
eleven-dimensional supergravity, it is elementary to dualize them to the appropriate IIB
frame.  One then obtains a solution in which the D1 and D5 branes are wrapped on a
common circle, $C$ .  To obtain a solution that is asymptotic to
$AdS_3 \times S^3 \times T^4$  one must also drop the constant terms in the
harmonic functions associated with the D1-brane and D5-brane charges
\refs{\entropy,\elvangtwo}.   In doing this, the circle, $C$, decompactifies
and becomes part of the $AdS_3$ \foot{One should not confuse this $AdS_3$
with that of the near-horizon limit of the supertube in M-theory:  They are different,
and the $AdS_3$ of the IIB theory emerges non-trivially via the T-dualities.}.

It is often useful to consider the D1-D5 field theory in a finite-sized ``box'' and one of
the simplest ways to do this is to restore the constants to the harmonic functions
so that the supergravity solution is asymptotically flat and the common circle, $C$,
has a radius, $R$.   At weak coupling, the perturbative string excitations must be quantized in
mass units of ${1 \over N_1 N_5 R}$ and so one expects the mass gap and the typical
energy gap between states to be of this order.   There are some issues as to whether
this approach is well-defined in the strict sense of the $AdS$/CFT correspondence (see below);
a more careful approach would be to introduce a UV cut-off in the radial direction of  $AdS_3$.
The effect of this is, once again, to introduce a scale
in the bulk.   More generally, anything that sets a finite scale for the spatial volume
of the field theory direction at infinity also sets a mass scale for that theory.

Three-charge solutions that are asymptotically $AdS_3 \times S^3 \times T^4$
also have additional, intermediate scales. For both black holes and black rings, there exists a scale  $r_p \sim \sqrt{Q_P}$ associated to the total momentum. This scale is set by the equal balance of the terms in the momentum harmonic function $Z_P \approx 1+Q_P/r^2$. For black rings there
are also scales set by the radius of the ring and by the dipole charges.

Since the $AdS$/CFT correspondence relates smooth, horizonless, asymptotically $AdS$
solutions  to states of the dual field theory, one can calculate, both in the bulk and on the
boundary, the spectrum of non-BPS excitations above a given BPS state, and try to identify
the boundary dual of a certain state by matching these spectra.  These calculations
have been very successful both for two-charge solutions
 \refs{\LuninJY,\LuninIZ}, and for simple three-charge solutions \GiustoIP.  This has
allowed precise matching of bulk solutions with boundary states.

A rougher way to estimate the non-BPS mass gaps in the spectrum of excitations
above a certain asymptotically-flat bulk solution is to consider the lowest energy
oscillations localized in the throat of this solution. The corresponding mass gap,
and indeed typical energy separation of states,  in the
holographic dual theory is then obtained by calculating the red-shifted energy of
these excitations at infinity in the asymptotically flat solution\foot{
Alternatively, one can work entirely with an asymptotically
$AdS$ solution, that is cut off at a large, but finite distance, $r = {1 \over \epsilon}$, and
impose appropriate boundary conditions on this surface  \FreedmanTZ.  The red-shifted bulk energy evaluated
at the cutoff $r = {1 \over \epsilon}$ can then be matched to the energy in the boundary theory placed in a box of size  ${1 \over \epsilon}$.}.  For the D1-D5-P system, introducing a
cut-off for evaluating the energy of excitations is not even necessary since the
bulk solution already contains a scale associated to the momentum charge $Q_P$. The
bulk energies redshifted to this scale correspond on the boundary to the ratio between
the energy of the excitations and the energy coming from the total momentum.

For the deep microstate
geometries,  the non-BPS excitations about these states  generically have a mass gap,
and typical energy separation between states, that vary inversely with the depth
of the throat in the bulk.  For the $U(1) \times U(1)$ invariant microstates, the spectrum
coming from the deepest possible throats matches the lowest bound on the mass
gap expected from the orbifold point description of the CFT, namely,
$E_0 = {1 \over N_1 N_5 R}$ \BenaKB. Although we have not checked this explicitly, we also
expect the mass gap of the deepest $U(1) \times U(1)$ black ring microstates to match the
mass gap expected from the orbifoled point description of black rings \entropy.

For microstates that do not have a  $U(1) \times U(1)$ invariance, we have the
``abysses''  in which  the throat can become arbitrarily deep as a function of moduli.
As the throat becomes deeper and deeper, all the excitations at the bottom of
the throat became lighter and lighter, and the field theory spectrum approaches what
looks like a continuum spectrum. Nevertheless, one does not expect this of a
quantum theory that is confined in a box, however large.   Since we are comparing
spectra at different values of the coupling constants, and one might  argue that
strong coupling effects do not modify the spectrum of $U(1) \times U(1)$
invariant configurations, but will modify the spectrum of the configurations with
less symmetry, and allow states whose energy separations are much smaller
than the expected  weak-coupling value.  However, for excitations above a given BPS state, these
energy separations will not become arbitrarily small if the size of the box is kept fixed.
Using condensed-matter language, when the mass of a very large number
of excitations goes to zero one approaches a quantum critical point, and one does not
expect to find quantum critical points in systems of finite size.

\subsec{A possible resolution:  Quantizing the moduli space}

The simplest and most straightforward resolution of this abyssal conundrum
would be to find a way of cutting off the throats of the scaling solutions at some
finite value.

As we have already noted, the modulus in our example must be extremely
finely tuned in order to obtain a very deep throat.  In string theory, and even in
supergravity, this moduli space will be quantized.  Indeed, one can try to quantize
it by considering the effective action for slow motions on the
moduli space and then apply quantum mechanics (where Planck's constant
will be related to ${1 \over N}$ effects). This will mean that there will be limits
on our ability to precisely localize GH points on the GH base metric and
thus localize the moduli sufficiently well to generate very deep throats.
The effectiveness of this will depend on the details of the correct physical
metric on the phase-space of the theory. If the metric on the moduli space comes
from the positions in the $\IR^3$ base of the GH geometry, then it
may well provide an effective and useful cut-off.  On the other hand,
the phase space measure may well be related to the complete physical metric
and it is hard to imagine how a quantization principle could cut-off
a throat that is several megaparsecs long.

Putting this more graphically, suppose that one is given a smooth, horizonless,
classical  solution of arbitrarily low curvature and $g_s$, and that has a length $10^{10}$
times larger than the maximum value consistent with mass gaps on the boundary
theory, it is very hard to imagine that quantum effects, which are intrinsically of
order $1/N$, will be able to destroy it.   Note that the puzzle is {\it not} about destroying
the  very large throats by throwing particles from infinity. For an arbitrarily deep throat
this can always be done, as any particle thrown in from infinity will eventually be
blue-shifted enough going down the throat to destroy it.   From the boundary perspective
the very deep throats correspond to very finely tuned superpositions of eigenstates,
and generic interactions with other states can easily destroy them.  The fact that
particles thrown in from infinity destroy the states does not make them physically
irrelevant\foot{In the same way in which the fact that one can throw elephants and
destroy a resonant cavity does not make the study of the modes of the cavity irrelevant.}
(though it will probably imply that non-BPS microstates will not have arbitrarily long
throats -- see the discussion below).  The puzzle comes from the {\it existence} and the
physics of BPS microstates of arbitrary long throats, and the fact that the excitations that live
at the bottom of the throat appear virtually massless from the point of view of the boundary theory.

Despite the concerns over ``quantizing away'' macroscopic geometries, there are
natural ways in which this might be realized. For example,
the angles on the base space could well be quantized because of the
quantization of angular momentum.   Given a certain bubbling solution, the value
of $J_R$ is determined entirely by the quantized fluxes on the bubbles \Jright,
and hence it is automatically quantized.  The angular momentum,  $J_L$, defined in \Jleft\
and \dipoles, not only  depends upon the quantized flux but also upon the
orientations of the bubbles.  Continuously varying an angle will therefore
generically yield non-integer values of $J_L$.   While this is certainly true of
the simple example considered in Section 5, and indeed will be true if one
varies a single modulus in almost any solution, one can easily construct solutions
in which there are moduli that do not change the total value of $J_L$.
For example, one could make a scaling solution with two identical bubbling black
rings on opposite planes; alternatively one could consider a $\ZZ_2$ symmetric bubbling black hole
(like the ``pincer'' solution studied in \BenaKB).   These configurations, which have $J_L=0$
because of symmetry, can still become arbitrarily deep and it seems unlikely that any quantization
coming from the {\it total} angular momentum could stop that.

On the other hand, one should also have some notion of a local quantization of the angular momentum.
This is because, in some circumstances, it is possible to separate different
components of a scaling solution in such a manner that it can be decomposed
into separate classical objects; the $J_L$ of each component must be quantized.
It thus seems plausible that the individual contributions, $\vec J_{L\, ij}$ in
\angmomflux, coming from each bubble should be quantized.  The
{\it magnitude} of the $\vec J_{L\, ij}$ is already quantized because of the quantization
of fluxes, and so the non-trivial content of this statement lies in the
quantization of the direction of $\vec J_{L\, ij}$.  In such a picture, the total
$J_L$ in \JLsimp\ would then be obtained by the standard rules for the addition of
spins in quantum mechanics.  If this picture were correct, then the ability to
classically orient an individual bubble would be limited by the inverse of
the magnitude of the flux that it carries.  The fine tuning needed to create
abysses would thus be limited.   We are currently examining whether these
angles are indeed quantized and how this limits the depth of throats.

Another possibility is that even if abysses exist, it does not make sense to talk
about their mass gaps because even a very small particle at the bottom of a throat could have
a large effect on the geometry and prevent the throat from becoming arbitrarily long.
An example of this can occur in the ``doubly-infinite''  $AdS_2$ throats that are encountered in  the near-horizon geometry of black holes and black rings. As discussed
in \MaldacenaUZ\foot{See equations (2.15) to  (2.17) of
that paper for more details.}, such infinite throats can by destroyed by the energy-momentum
tensor coming from a very small perturbation.

Since the throats of our solutions are capped, the metric near the cap is no longer of
$AdS \times S$ form. Therefore, extending the calculation of \MaldacenaUZ\ to our solutions is not straightforward. If we naively assume equation (2.16) of \MaldacenaUZ\ captures the
essential physics, one can extend that analysis to our case. We find that a
non-trivial energy-momentum tensor can be accommodated on top of our solutions
 provided the sphere shrinks to zero
size at the cap. Fortunately, this is already happening even in the smooth
BPS solutions, and is indeed a necessary feature of all the capped microstates.
Hence, the obvious extension of the argument in \MaldacenaUZ\ does not rule
out abyssal throats. It would be very interesting to see if one can construct an
argument in a similar spirit that would cut off an abyss.

Summarizing this sub-section, it appears difficult to logically exclude the existence
of  some quantum mechanism that limits the depth of a throat.  On the other hand, if
this were to happen, this would be a rather remarkable first example of quantum
effects destroying a very large portion of a smooth, horizonless, low-curvature,
asymptotically-flat classical geometry.

\subsec{A possible resolution:  Stringent constraints on the duality}

Another possible resolution of the problem is to take the more ``stringent'' view
that the  $AdS$/CFT correspondence only relates field theories in an infinite volume to
asymptotically $AdS$ solutions without a cutoff. In this context calculations of
mass gaps, times of flights, or energy spectra are, at best, of limited validity and, at worst,
meaningless.   From this perspective, the only thing one can meaningfully compute in
the bulk are N-point functions.   Indeed, by computing one-point functions in certain
two-charge geometries and relating them to vev's in the boundary theory it is possible
to obtain a very precise mapping between bulk solutions and their dual boundary states \KanitscheiderWQ, without appealing to spectra and mass-gaps.

This view poses the opposite problem:  Why were mass-gap
calculations in the D1-D5 system in a finite box so successful? One possible
answer is that all these calculations were done for $U(1)\times U(1)$ invariant
microstates, and the extra symmetry ``protects''  the calculations done on the
two sides of the duality, even if the duality is not strictly valid.  Conversely,
the microstates  that do not have a $U(1)\times U(1)$ invariance are not protected, and
there is no reason why the calculations done in two inequivalent theories should agree.
This answer is also consistent with the fact that mass-gaps computed in $U(1) \times U(1)$
invariant three-charge microstates cannot be less than the smallest mass gap expected
from the free (orbifold point) description of the CFT \BenaKB.  If this view is correct then
one needs to understand what this protection mechanism is and why it works and
why it fails.

There is also another rather puzzling feature of this perspective:  While the D1-D5
system does not have a scale, the  D1-D5-P system does have a scale set by $Q_P$.
Even if it does not make sense to  talk about the  energy of excitations of arbitrarily
deep throats by themselves, it does  make sense to talk about the ratio between this
excitation energy and the energy  coming from the total momentum, $Q_P$.
We therefore find that the energies of excitations in an abyss are going to zero
compared to the energy of the momentum excitations that are ultimately responsible for
making the classical horizon area macroscopic.  It would therefore seem that
one could store very large amounts of entropy in such massless excitations.

\newsec{Entropy elevators and non-BPS microstates.}

We have thus two distinct, though logically possible outcomes, both
of which are physically unexpected and both of which hint at tantalizing new
phenomena.   If abysses are cut off by quantum  effects then these quantum
effects can remove macroscopic portions of a low-curvature asymptotically-flat
solution, and if the abysses are not cut off, then we appear to have a quantum critical point.

In trying to understand both of these possibilities we have found it useful
to think about the effects of ``entropy elevators.''  The  idea is to consider a
small sub-system of non-BPS excitations and then adiabatically lower that sub-system
into very deep throats so that the energy is red-shifted to
the value determined by the depth of the throat and yet the entropy in the non-BPS
excitations remains constant.  There are two ways that one could imagine controlling
such an elevator:  Either by lowering the elevator using a massless cable or, as we
prefer here, constructing the non-BPS excitations in a ``shallow cap'' and then
adiabatically changing the modulus so that the cap descends to the bottom
of a deep throat.  If the moduli space is quantized then the elevator is only allowed
to go to discrete floors and there is a lowest possible floor, but in an abyss there
is no lower limit and all the excitations of that sub-system will become massless in the
limit when the length of the throat approaches infinity. Hence, at the critical point the
system has new massless modes.

For a given smooth cap there will be a limit on the size of the non-BPS sub-system:
It must not significantly alter the  physics of the cap and radically modify the elevator.
In the limit when the throat is infinite, the extra mass will be zero and the state will,
once again be (arbitrarily close to) BPS, and yet there will still be entropy in this
sub-system.  The solution will also be (arbitrarily close to) the classical BPS black hole.
Hence, it is possible that one can only use microstate geometries to account for the
entropy of the BPS black hole by considering  all the  throats that can act as
``entropy elevators'' that carry massive sub-systems of finite entropy to an infinite throat
depth, where their mass becomes zero.

The entropy that a certain ``elevator'' can carry is limited by the requirement that
the non-BPS sub-system added on does not destroy the solution.
Note that this requirement has nothing to do with the energy seen from infinity,
but rather with the effect of the sub-system on the bubbles that form the cap of
the solution.  Whether the sub-system destroys a certain cap, or not, has
nothing to do with the length of the throat at whose end the cap is.  The sub-system
should only care about the local geometry of the cap and its presence
should not limit the ability of the elevator to descend.  The only effect of
the throat is to make the energy of the sub-system as seen from infinity larger or smaller.
Thus, for every cap we can associate a maximal ``local'' energy $E_c$ that is the
maximal energy that does not destroy it, and a certain entropy $S_c$.
The mass above the BPS bound  as seen from infinity is
$E_{\infty} = E_c \sqrt{g_{00}^{min}}$ where ${g_{00}^{min}}$
is the value of ${g_{00}}$ at the bottom of the throat. As the throat length
approaches infinity, the elevator associated with each deep throat contributes
with $S_c$ to the entropy of the black hole. It is tempting to conjecture that the
entropy of the BPS black holes comes entirely from these entropy elevators\foot{This would imply that
the entropy of the D1-D5-P CFT at strong coupling has ``accumulation points,'' corresponding to the
abysses.}.

The idea of lowering boxes containing entropy into black holes and studying the
entropy in the process is not new in General Relativity  and has led to apparent
paradoxes and beautiful resolutions.  See, for example,  \refs{\UnruhIC,\UnruhIR} and for
recent work see also \MarolfAY. However, entropy elevators have two very different
features. The first is that the box is lowered on top of a horizonless, BPS solution, and there
is no Hawking radiation from the horizon to keep the box in equilibrium, or to allow the
creation of box - anti-box pairs. The second is that the entropy in the elevators never
goes into the entropy of a black hole: As the elevators descend, the solution always
remains horizonless.  Indeed, we will see below that as an elevator carrying a box of a
certain ``local'' energy, $E$, descends, the energy as seen from infinity decreases, and
the horizon of the corresponding black hole also descends at the same rate.

The idea of entropy elevators also has some interesting consequences for
near-BPS black holes.   These black holes have a finite
throat depth, set by the non-extremality parameter. In the elevator picture, to create a finite amount of non-extremality one must add a finite amount of energy, $\Delta E$, above the BPS bound.  To create this amount of energy at infinity by putting a non-BPS sub-system on an elevator means that the
 amount of energy on the elevator must be $E_{local} = \Delta E/\sqrt{g_{00}}$.
At a certain depth this energy will exceed the energy, $E_c$, needed to destroy
the cap.  Thus there is a limit to which the entropy elevator can descend for
a given amount of non-extremality.

If the entropy elevators can be used to create near-BPS black hole microstates, the depth of the
entropy elevators that carry most of the entropy
should match the depth of the horizon of the near-BPS black hole.
While a perfect matching of these two quantities is not possible without constructing
the solutions corresponding to the elevators, one can check that the depth of the
elevators and the ``depth'' of the horizon scale in the same way with the energy
above extremality, which is a rather non-trivial check.

Indeed, given a certain energy, $\Delta E$, above the BPS bound, one can construct
shallow entropy elevators, that have a sub-system of energy a few times $\Delta E$, as
well as deeper elevators, that have a sub-system of energy $E_{local} =
\Delta E/\sqrt{g_{00}^{\rm bottom}}$. Clearly, the deeper elevators will carry a
bigger system, and will have more entropy. If we make $E_{local}$ bigger than
the maximal energy a cap can support, $E_c$, then the elevator will be destroyed.
Hence, most entropy will come from the elevators of depth corresponding to
\eqn\gdeep{g_{00}^{\rm bottom} = (Z_1 Z_2 Z_3)^{-2/3} = {(\Delta E)^2 \over E_c^2}. }
We can compare the depth of these elevators to the depth of the horizon of a
near-BPS black hole. The easiest measure of the depth of that throat are the values of
the three harmonic functions at the horizon, which in the near-BPS limit are given by:
\eqn\zbhhor{Z_i \approx{ Q_i \over \Delta E}}
Hence, the depth of the elevators and the depth of the near-BPS black hole horizon have the same
dependence on $\Delta E$. This is a necessary feature if elevators are to give the microstates of
the non-extremal black holes, and its confirmation is encouraging. It would be interesting
to analyze in more detail the amount of energy, $E_c$, that a certain cap can carry,
and see if its dependence on the charges also matches that predicted by
equations \gdeep\ and \zbhhor.

Hence, if this idea of entropy elevators is correct,  one should think about elevators
that descend to an infinite depth as giving the entropy of the extremal black holes,
and the elevators that descend to a finite depth as giving the entropy of the
non-extremal black holes.

\bigskip
\bigskip

\noindent{\bf Acknowledgments:~}
We would like to thank Davide Gaiotto, Eric Gimon, Juan Maldacena, Samir Mathur, Radu Roiban, Ashoke Sen,
Kostas Skenderis, Marika Taylor and Xi Yin for interesting discussions. NPW would like to thank the SPhT, CEA-Saclay for hospitality while this work was completed.
The work of NPW and CWW is supported
in part by the DOE grant DE-FG03-84ER-40168.  The work of IB was supported in
part by the {\it Dir\'ection des Sciences de la Mati\`ere}
of the {\it Commissariat \`a l'En\'ergie Atomique} of France and by the ANR grant
BLAN06-3-137168.

\appendix{A}{Verifying the absence of closed time-like curves}

To check that there are no CTC's we must verify the standard set of conditions:
\eqn\noCTCs{ V \, Z_I ~ \ge~ 0 \,, \qquad  {\cal Q} ~\equiv~ Z_1\,Z_2 \, Z_3 \, V ~-~
\mu^2 \, V^2 ~\ge~ 0 \,.}
We have extensively examined these conditions numerically and found
that they are satisfied for a broad sample of the scaling solutions given in
Table 1.  While the conditions, \noCTCs, are necessary, there is an additional dangerous
term from the angular terms in the direction of the base metric:
\eqn\baseproj{ (\, Z_1\, Z_2\, Z_3\, )^{1/3}\, V\, \Big(\, dx^2 + dy^2 + dz^2 ~-~
{\omega^2 \over \cal{Q}}\, \Big) ~\equiv~ (\, Z_1\, Z_2\, Z_3\,
)^{1/3}\, V\, \bar g_{\mu\nu}\,dx^{\mu}\,dx^{\nu}\,. }
Since the positivity of the coefficient, $(Z_1\, Z_2\, Z_3\, )^{1/3}\, V$, has already been
verified in checking  \noCTCs,  it remains to verify  that  $\bar g_{\mu\nu}$ has no
negative-norm vectors.  The  lowest eigenvalue of $\bar g_{\mu\nu}$ must therefore
be non-negative:
\eqn\noCTCfourth{ 1 ~-~ {|\omega|^2 \over {\cal Q}} ~\ge~ 0\,.}
We checked this condition by cutting a slice near the ring and numerically
evaluating this condition on this slice. We then moved this slice around
to check the potentially dangerous region. In Figure 4, we show a particular
slice that covers all three charges.  We found that \noCTCfourth\ was globally satisfied,
and $\cQ -  |\omega|^2$ was globally positive.  Not
only is this a more stringent condition that the last condition in \noCTCs, but,
as was noted in \BerglundVB, this means that the complete metric is stably causal
and that the coordinate, $t$, provides a global time function.

\goodbreak\midinsert
\vskip .2cm
\centerline{ {\epsfxsize 2.4in\epsfbox{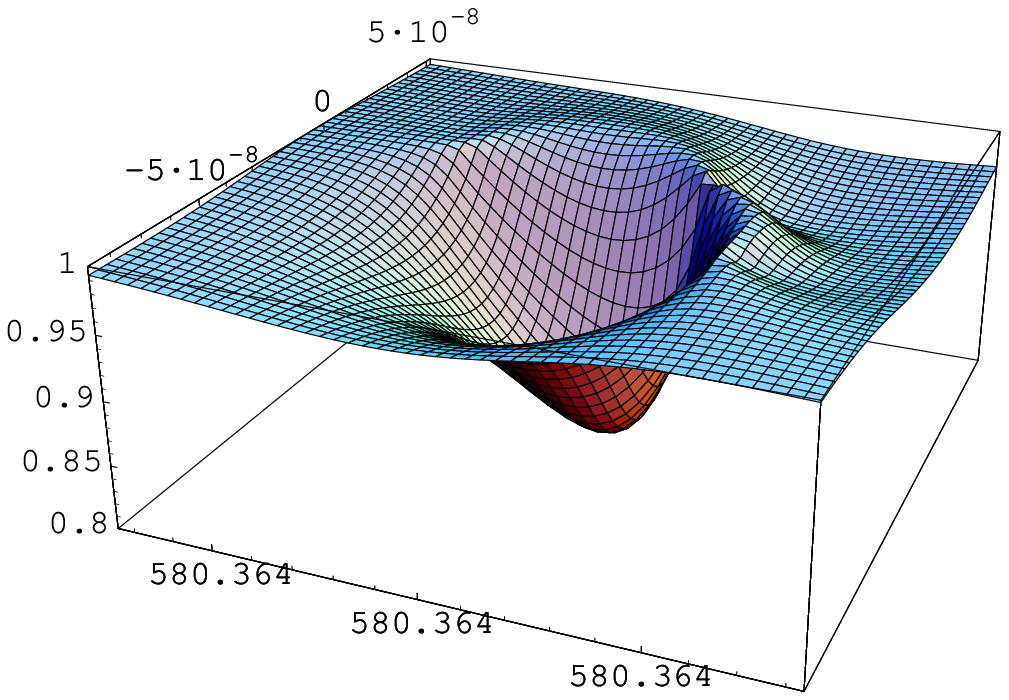}}
\hskip 0.5in {\epsfxsize 2.4in\epsfbox{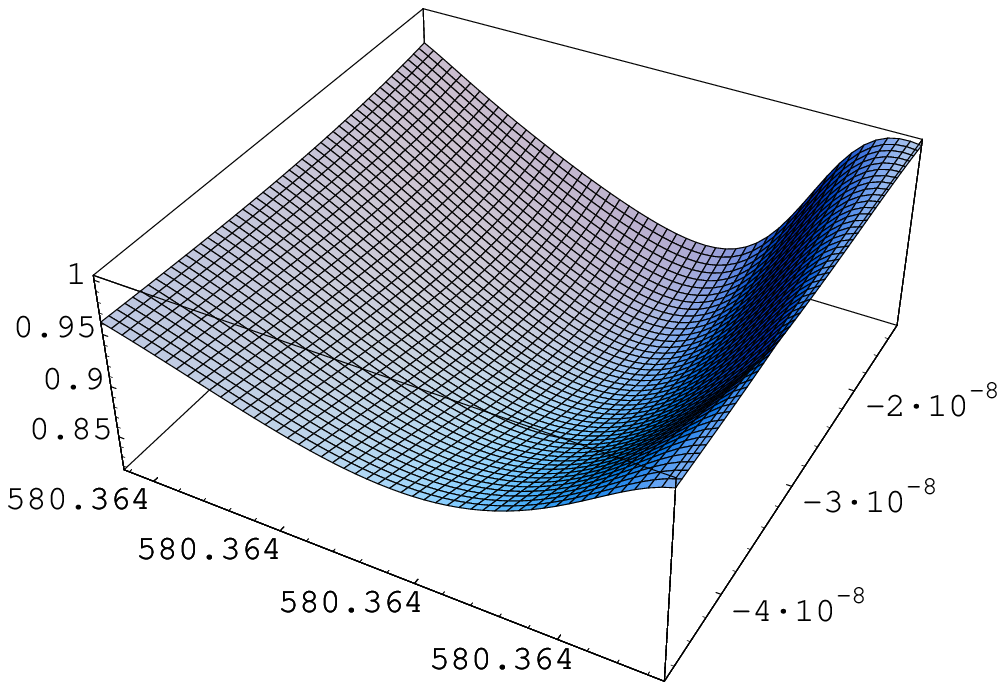}}}
\vskip 0.1cm
\leftskip
2pc \rightskip 2pc\noindent{\ninepoint\sl \baselineskip=8pt {\bf Fig.~4}:
These two graphs evaluate the condition \noCTCfourth$\,$ on the plane
where all three charges are located for a particular triangle solution for which
the three distances are $\sim 10^{-8}$. The first graph covers all three
charges and the second graph shows a ``zoom-in'' at the bottom of the valley
in the first graph. }
\endinsert

\listrefs
\vfill
\eject
\end